\documentclass[a4paper,UKenglish]{lipics-v2016}

\usepackage{color}
\newcommand{\annotate}[1]{{#1}}

\usepackage{microtype}
\usepackage{tikz}
\newcommand*\circled[1]{\tikz[baseline=(char.base)]{
	            \node[shape=circle,draw,inner sep=2pt] (char) {#1};}}
\usepackage{floatrow}
\newfloatcommand{capbtabbox}{table}[][\FBwidth]

\title{End-to-end Analysis and Design of a Drone Flight Controller}
\titlerunning{End-to-end Analysis and Design of a Drone Flight Controller} 

\author[1]{Zhuoqun Cheng}
\author[1]{Richard West}
\author[1]{Craig Einstein}
\affil[1]{Boston University, USA\\
  \texttt{czq@cs.bu.edu}\\ 
  \texttt{richwest@cs.bu.edu}\\ 
  \texttt{einstein@cs.bu.edu}}
\authorrunning{Z.\,Cheng, R.\,West and C.\,Einstein} 


\ArticleNo{}

\begin{document}

\maketitle

\begin{abstract}
Timing guarantees are crucial to cyber-physical applications that must bound
the end-to-end delay between sensing, processing and actuation. For example,
in a flight controller for a multirotor drone, the data from a gyro or
inertial sensor must be gathered and processed to determine the attitude of
the aircraft. Sensor data fusion is followed by control decisions that adjust
the flight of a drone by altering motor speeds. If the processing pipeline
between sensor input and actuation is not bounded, the drone will lose control
and possibly fail to maintain flight.

Motivated by the implementation of a multithreaded drone flight controller on
the Quest RTOS, we develop a composable pipe model based on the system's task,
scheduling and communication abstractions. This pipe model is used to analyze
two semantics of end-to-end time: reaction time and freshness time. We also
argue that end-to-end timing properties should be factored in at the early
stage of application design.  Thus, we provide a mathematical framework to
derive feasible task periods that satisfy both a given set of end-to-end
timing constraints and the schedulability requirement.  We demonstrate the
applicability of our design approach by using it to port the Cleanflight
flight controller firmware to Quest on the Intel Aero board. Experiments show
that Cleanflight ported to Quest is able to achieve end-to-end latencies
within the predicted time bounds derived by analysis.
\end{abstract}

\section{Introduction}
\label{sect:intro}
Over the past few years, commercial and hobbyist multirotor drones have been
rapidly growing in popularity. The fast development of drone technology
enables an ever widening set of applications, including aerial
photography~\cite{DJI}, package delivery~\cite{AmazonAir}, and search and
rescue~\cite{Valavanis:book08,BBC:SearchRescue}. One of the most commonly used
control boards for drones in use today is the STM32 family of systems-on-chip
(SoCs), which are based on the ARM Cortex M-series processors, and include
integrated inertial sensors such as a gyroscope, accelerometer and
magnetometer.

While many of the existing flight control boards are perfectly adequate for
drones operated via human-assisted radio control, they fall short of the
processing capabilities needed for fully autonomous operation. For this
reason, we are developing a new approach to building autonomous drones using
emerging multicore platforms such as the Intel Aero board, Qualcomm Snapdragon
flight development board~\cite{snapdragon_flight_kit}, and Nvidia's
Jetson~\cite{nvidia_jetson}. All these boards offer multiple processing cores
and integrated graphics processing capabilities, making them capable of
mission tasks that would be impossible on simpler hardware.

Our first step to building autonomous drones has so far involved a
reimplementation of the popular racing drone flight control firmware called
Cleanflight~\cite{cleanflight} on the Intel Aero board. We have also ported
our in-house Quest real-time OS (RTOS)~\cite{quest} to the Aero board, to
efficiently and predictably manage the multiple cores and I/O complexity.  Our
reimplementation of Cleanflight refactors the original single-threaded code
running directly as firmware on STM32 SoCs into a multi-threaded application
running on Quest. The decoupling of software components into separate threads
improves the modularity of Cleanflight, and provides the capability for
parallel task execution on platforms with multiple cores. Flight management
tasks are then able to leverage the availability of increased compute
resources, potentially improving the controllability of a drone. Similarly,
the cleaner interfaces between software components eases our future plans to
extend Cleanflight with advanced features such as camera data processing,
object detection and avoidance, and simultaneous localization and mapping
(SLAM) necessary for autonomous flight management.

The original Cleanflight code has a series of tasks that are executed as loops
with predefined frequencies. These frequencies are based on a combination of
the capabilities of the hardware and the experiences of drone developers.
However, a multithreaded Cleanflight is subject to extra overheads and
uncertainty, due to scheduling and inter-task communication. It is critical to
ensure the timing correctness of Cleanflight. For example, if a gyroscope
reading fails to correctly influence a change in motor (and, hence, rotor)
speed within a specific time bound, the drone might not be able to stabilize.

Cleanflight is typical of many applications that process sensor inputs and
require time-bounded changes to actuators. With the recent development of
multi-sensor data fusion algorithms~\cite{mahony, madgwick, sensor_fusion} and
ever-increasing availability of open source hardware, multi-sensor/actuator
cyber-physical applications are leading to a revolution in areas such as 3D
printing, drones, robotics, driverless cars, and intelligent home automation
systems.  For these applications, it is essential to guarantee two types of
end-to-end timing requirements: 1) the maximum time it takes for an input
sensor reading to flow through the whole system to eventually affect an
actuator output, and 2) the maximum time within which an input sensor reading
remains influential on output actuator commands.

While the real-time community has developed valuable approaches to scheduling
and response time analysis of tasks~\cite{response_time}, end-to-end timing
analysis has received only limited attention. Most of the prior work has
originated from real-time network communication
research~\cite{AFDX,Fisher,Fohler,Goddard,CAN_response,CAN_response2}, and is
based on event-triggered communication with FIFO-based buffers.  In a drone
flight control program, however, single register-based buffers and periodic
sampling are more common. This paper, therefore, presents the end-to-end
timing analysis of a drone flight controller based on a combination of the
periodic-sampling task model, the register-based inter-task communication
model and the Quest RTOS's scheduling model. We also show how the derived
worst case end-to-end communication time can be, in turn, used to guide the
design of applications.

Contributions of this paper include: 1) the proposal of a composable pipe
model to capture the timing characteristics of end-to-end communication in the
Quest RTOS; 2) a demonstration of how to derive task periods from given
end-to-end timing constraints in the application design stage; 3) the
re-implemention and evaluation of the Cleanflight flight controller on the
Intel Aero board.

The rest of the paper is organized as follows: Section~\ref{sect:model}
provides background on Cleanflight and the corresponding task, scheduling and
communication models adopted throughout the
paper. Section~\ref{sect:e2e_analysis} describes the end-to-end timing
analysis of our proposed composable pipe model. Section~\ref{sect:e2e_design}
shows how the end-to-end time is leveraged in the application design stage,
while Section~\ref{sect:eval} details the re-implementation and evaluation of
Cleanflight on the Aero board. Related work is discussed in
Section~\ref{sect:related}, followed by conclusions and future work in
Section~\ref{sect:conclusion}.

\section{Execution Model}
\label{sect:model}
In this section, we first provide an overview of the application that
motivated this work. Secondly, we describe the application design model from
three perspectives: \textit{i.e.,} the task model, the scheduling model and
the communication model.

\subsection{Application and System Overview}
This paper is motivated by our objective to implement an autonomous flight
management system for multirotor drones. An autonomous drone is one that is
able to reason about and adapt to changes in its surroundings, while
accomplishing mission objectives without remote assistance from a human being
(once its objectives are established). As part of this effort, we have
undertaken a port of the Cleanflight firmware from a traditional single-core
STM32-based system-on-chip (SoC) to the Intel Aero compute board. The Aero
board has a quadcore x86 Atom x7-Z8750 processor, 4GB RAM, an integrated GPU,
inertial measurement sensors and 3D camera connectivity. This makes it capable
of flight management tasks (e.g., package delivery, aerial photography, search
and rescue) that would be impossible with a less powerful single-core ARM
Cortex M3 or M4 found in most STM32 SoCs.

Cleanflight is targeted at racing drones, which are operated by humans using
radio control. The core software components of Cleanflight consist of sensor
and actuator drivers, a PID controller, the Mahony Attitude and Heading
Reference (AHRS~\footnote{The attitude is the orientation of the drone
relative to a reference frame such as earth.}) algorithm, various
communication stacks, and a logging system. Runtime entities of those
components are called tasks.  There are 31 tasks in total, of which more than
half are optional.  The essential ones are listed in
Table~\ref{tab:tasklist}. Tasks are scheduled from highest to lowest dynamic
priority, calculated as a function of a task's static priority and the time
since it was last executed.

\begin{table}[!ht]
\begin{small}
\begin{center}
\begin{tabular} {cccc}
	{\bf Task Name} & {\bf Period ($\mu$s)} & {\bf Static Priority} & {\bf Description}\\ \hline
	\texttt{System} & \texttt{100000} & \texttt{Medium-High} & \texttt{Check system utilization}\\ \hline
	\texttt{Battery Alert} & \texttt{200000} & \texttt{Medium} & \texttt{Alarm battery runout}\\ \hline
	\texttt{Battery Voltage} & \texttt{20000} & \texttt{Medium} & \texttt{Update battery voltage reading}\\ \hline
	\texttt{Battery Current} & \texttt{20000} & \texttt{Medium} & \texttt{Update battery current reading}\\ \hline
	\texttt{Gyro} & \texttt{1000} & \texttt{RealTime} & \texttt {Update gyro readings} \\ \hline
	\texttt{PID} & \texttt{1000} & \texttt{RealTime} & \texttt {Perform PID-based motor control} \\ \hline
	\texttt{Accelerometer} & \texttt{1000} & \texttt{Medium} & \texttt {Update accelerometer readings} \\ \hline
	\texttt{IMU Attitude} & \texttt{10000} & \texttt{Medium} & \texttt {Calculate attitude}\\ \hline
	\texttt{RC Receiver} & \texttt{20000} & \texttt{High} & \texttt {Process RC commands} \\ \hline
	\texttt{Serial} & \texttt{10000} & \texttt{Low} & \texttt {Serial communication} \\ \hline
	\texttt{Magnetometer} & \texttt{100000} & \texttt{Low} & \texttt {Update magnetometer readings} \\
\end{tabular}
\end{center}
\end{small}
~\vspace{0.02in}
\caption{List of Cleanflight Tasks}
\label{tab:tasklist}
\vspace{-0.3in}
\end{table}

Our port of Cleanflight to the Aero board runs on our in-house Quest RTOS.  We
have developed Quest drivers for SPI, I2C, GPIO, UART, and inertial sensors on
this and other similar boards. Quest is an SMP system, providing both user
and kernel level threads, as well as threaded interrupt handlers that are
scheduled by a time-budgeted virtual CPU (VCPU) scheduler detailed in
Section~\ref{sect:sched}.

\subsection{Task Model}
\label{sect:task}
We model the flight controller program as a set of real-time periodic tasks
$\{\tau^{1}, \tau^{2}, \cdots, \tau^{n}\}$. Each task $\tau^{j}$ is
characterized by its worst case execution time $e^{j}$ and period $T^{j}$.
$e^{j}$ and $T^{j}$ are determined during the design stage and are fixed
at runtime. $e^{j}$ is usually profiled off-line under the worst case
execution condition. Deciding the value of $T^{j}$ is a challenging process,
which is the major topic of this paper. It mainly depends on the end-to-end
latency constraints and the schedulability test. All the periodic tasks
are implemented using Quest's user level threads. In this paper, we use
term \verb+thread+ and \verb+task+ interchangeably.

Apart from user level threads, there are kernel threads dedicated to I/O
interrupts, which originate primarily from the SPI and I2C bus in
Cleanflight. Quest executes interrupts in a deferrable thread context, having
a corresponding time budget. This way, the handling of an interrupt does not
steal CPU cycles from a currently running, potentially time-critical task.

\subsection{Scheduling Model}
\label{sect:sched}
Threads in Quest are scheduled by a two-level scheduling hierarchy, with
threads mapped to virtual CPUs (VCPUs) that are mapped to physical CPUs.  Each
VCPU is specified a processor capacity reserve~\cite {processorcapacity}
consisting of a budget capacity, $C$, and period, $T$. The value of $C$ and
$T$ are determined by the $e$ and $T$ of the task mapped to the VCPU. A VCPU
is required to receive at least C units of execution time every T time units
when it is runnable, as long as a schedulability test~\cite {Lehoczky:89} is
passed when creating new VCPUs. This way, Quest's scheduling subsystem
guarantees temporal isolation between threads in the runtime environment.

Conventional periodic tasks are assigned to Main VCPUs, which are implemented
as Sporadic Servers~\cite{spruntsporadic} and scheduled using Rate-Monotonic
Scheduling (RMS)~\cite{RMS}. The VCPU with the smallest period has the highest
priority. Instead of using the Sporadic Server model for both main tasks and
bottom half threads, special I/O VCPUs are created for threaded interrupt
handlers. Each I/O VCPU operates as a Priority Inheritance Bandwidth
preserving Server (PIBS)~\cite{quest-vcpu}. A PIBS uses a single replenishment
to avoid fragmentation of replenishment list budgets caused by short-lived
interrupt service routines (ISRs).  By using PIBS for interrupt threads, the
scheduling overheads from context switching and timer reprogramming are
reduced~\cite {MissimerPIBS}.


\subsection{Communication Model}
\label{sect:async_comm}

Control flow within the flight controller is influenced by the path of data,
which originates from sensory inputs and ends with actuation. Inputs include
inertial sensors, optional cameras and GPS devices, while actuators include
motors that affect rotor speeds and the attitude of the drone.  Data flow
involves a pipeline of communicating tasks, leading to a communication model
characterized by: (1) the interarrival times of tasks in the pipeline, (2)
inter-task buffering, and (3) the tasks' access pattern to communication
buffers.

\textbf{Periodic vs. Aperiodic Tasks.} Aperiodic tasks have irregular
interarrival times, influenced by the arrival of data. Periodic tasks have
fixed interarrival times and operate on whatever data is available at the time
of their execution. A periodic task implements asynchronous communication by
not blocking to await the arrival of new data~\cite{Feiertag2008}.

\textbf{Register-based vs. FIFO-based Communication.} A FIFO-based shared  
buffer is used in scenarios where {\em data history} is an important
factor. However, in a flight controller, {\em data freshness} outweighs the
preservation of the full history of all sampled data. For example, the motor
commands should always be computed from the latest sensor data and any stale
data should be discarded.  Moreover, FIFO-based communication results in
loosely synchronous communication: the producer is suspended when the FIFO
buffer is full and the consumer is suspended when the buffer is
empty. Register-based communication achieves fully asynchronous communication
between two communicating parties using Simpson's four-slot
algorithm~\cite{4slots}.

\textbf{Implicit vs. Explicit Communication.} Explicit communication allows 
access to shared data at any time point during a task's execution. This might
lead to data inconsistency in the presence of task preemption.  A task that
reads the same shared data at the beginning and the end of its execution might
see two different values, if it is preempted between the two reads by another
task that changes the value of the shared data.  Conversely, the implicit
communication model~\cite{ecrts2017comm} essentially follows a
read-before-execute paradigm to avoid data inconsistency. It mandates a task
to make a local copy of the shared data at the beginning of its execution and
to work on that copy throughout its execution. 

This paper assumes a periodic task model, as this simplifies timing
analysis. Applications such as Cleanflight implement periodic tasks
to sample data and perform control operations. Our system also adopts
register-based, implicit communication for data freshness and consistency.

\section{End-to-end Communication Timing Analysis}
\label{sect:e2e_analysis}

In this section, we first distinguish two different timing semantics for
end-to-end communication, which will be used as the basis for separate timing
analyses. Secondly, we develop a composable pipe model for communication,
which is derived from separate latencies that influence end-to-end delay.
Lastly, we use the pipe model to derive the worst case end-to-end
communication time under various situations.

\subsection{Semantics of End-to-end Time}
\label{sect:semantic}

\begin{figure}[!ht]
	\begin{floatrow}
	  \capbtabbox{%
	    \begin{tabular}{|c|c|c|} \hline
	      & \texttt{Gyro Period} & \texttt{Motor Period} \\ \hline
	      \texttt{Case 1} & 10 ms & 1 ms\\ \hline
	      \texttt{Case 2} & 1 ms & 10 ms\\ \hline
	    \end{tabular}
	  }{%
	    \caption{Example Periods}%
	    \label{tab:example_case}
	  }
          
	  \ffigbox{%
	    \includegraphics[scale=.35]{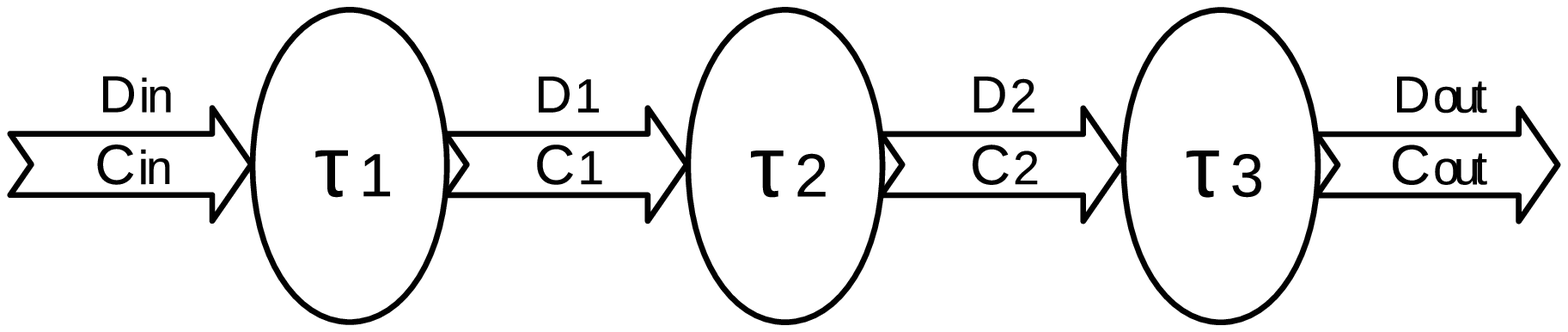}
	  }{%
	    \caption{Task Chain}
	    \label{fig:task_chain1}
	  }
	\end{floatrow}
\end{figure}

To understanding the meaning of end-to-end time consider the following two
constraints for a flight controller:
\begin{itemize}
 \item Constraint 1: a {\em change} in motor speed must be within 2 ms of the
   gyro sensor reading that caused the change.
 \item Constraint 2: an {\em update} to a gyro sensor value must be within 2
   ms of the corresponding update in motor speed.
\end{itemize}
The values before and after a {\em change} differ, whereas they may stay
the same before and after an {\em update}. These semantics lead to two
different constraints. To appreciate the difference, imagine the two cases in
Table~\ref{tab:example_case}.  In Case 1, the task that reads the gyro runs
every 10 ms and the one that controls the motors runs every 1 ms. Case 1 is
guaranteed to meet Constraint 1 because the motor task runs more than once
within 2 ms, no matter whether the gyro reading changes. However, it fails
Constraint 2 frequently as the gyro task is not likely to run even once in an
interval of 2 ms.  Conversely, Case 2 is guaranteed to meet Constraint 2 but
fails Constraint 1 frequently.

This example demonstrates the difference between the two semantics of
end-to-end time, which leads to the following formal definitions:
\begin{itemize}
\item \verb+Reaction time+ is the time it takes for input data to flow through
  the system\annotate{, and is affected by the period of each
    consumer in a pipeline}. A reaction timing constraint bounds the time
  interval between a sensor input and the {\em first corresponding} actuator
  output.
\item \verb+Freshness time+ is the time within which an instance of the input
  data has influence on the system\annotate{, and is affected by the period
  of each producer in a pipeline}. A freshness timing
  constraint bounds the time interval between a sensor input and
  the {\em last corresponding} actuator output. 

\end{itemize}
Constraint 1, above, is a constraint on reaction time, while Constraint 2 is
on freshness time. We perform analysis of the two semantics of time in
Section~\ref{sect:reaction} and~\ref{sect:freshness}, respectively.

\subsection{Latency Contributors} \label{sect:latency_contr}
The end-to-end communication delay is influenced by several factors, which we
will identify as part of our analysis. To begin, we first consider the
end-to-end communication pipeline illustrated as a task chain in
Figure~\ref{fig:task_chain1}. Task $\tau_1$ reads input data $D_{in}$ over
channel $C_{in}$, processes it and produces data $D_1$. Task $\tau_2$ reads
$D_1$ and produces $D_2$, and $\tau_3$ eventually writes output $D_{out}$ to
channel $C_{out}$ after reading and processing $D_2$.

Each task handles data in three stages, \textit{i.e.}, read, process and
write.  The end-to-end time should sum the latency of each stage in the task
chain. Due to the asynchrony of communication, however, we also need to
consider one less obvious latency, which is the waiting time it takes for an
intermediate output to be read in as input, by the succeeding task in the
chain.  In summary, the latency contributors are classified as follows:
\begin{itemize}
	\item \verb+Processing latency+ represents the time it takes for a
          task to translate a raw input to a processed output. The actual
          processing latency depends not only on the absolute processing time
          of a task without interruption, but also on the service constraints 
          (\textit{i.e.}, CPU budget and period of the VCPU) associated with
          the task.
	\item \verb+Communication latency+ represents the time to transfer
          data over a channel. The transfer data size, bandwidth and
          propagation delay of the communication channel, and the software
          overheads of the communication protocol all contribute to the overall
          latency. Since our communication model is asynchronous and
          register-based as described in Section~\ref{sect:model}, queuing
          latency is not a concern of this work.
	\item \verb+Scheduling latency+ represents the time interval between
          the arrival of data on a channel from a sending task and when the
          receiving task begins reading that data. The scheduling latency
          depends on the order of execution of tasks in the system, and
          therefore has significant influence on the end-to-end communication
          delay.
\end{itemize}

\subsection{The Composable Pipe Model}
\begin{figure}[!ht]
  \centering
  \includegraphics[scale=.5]{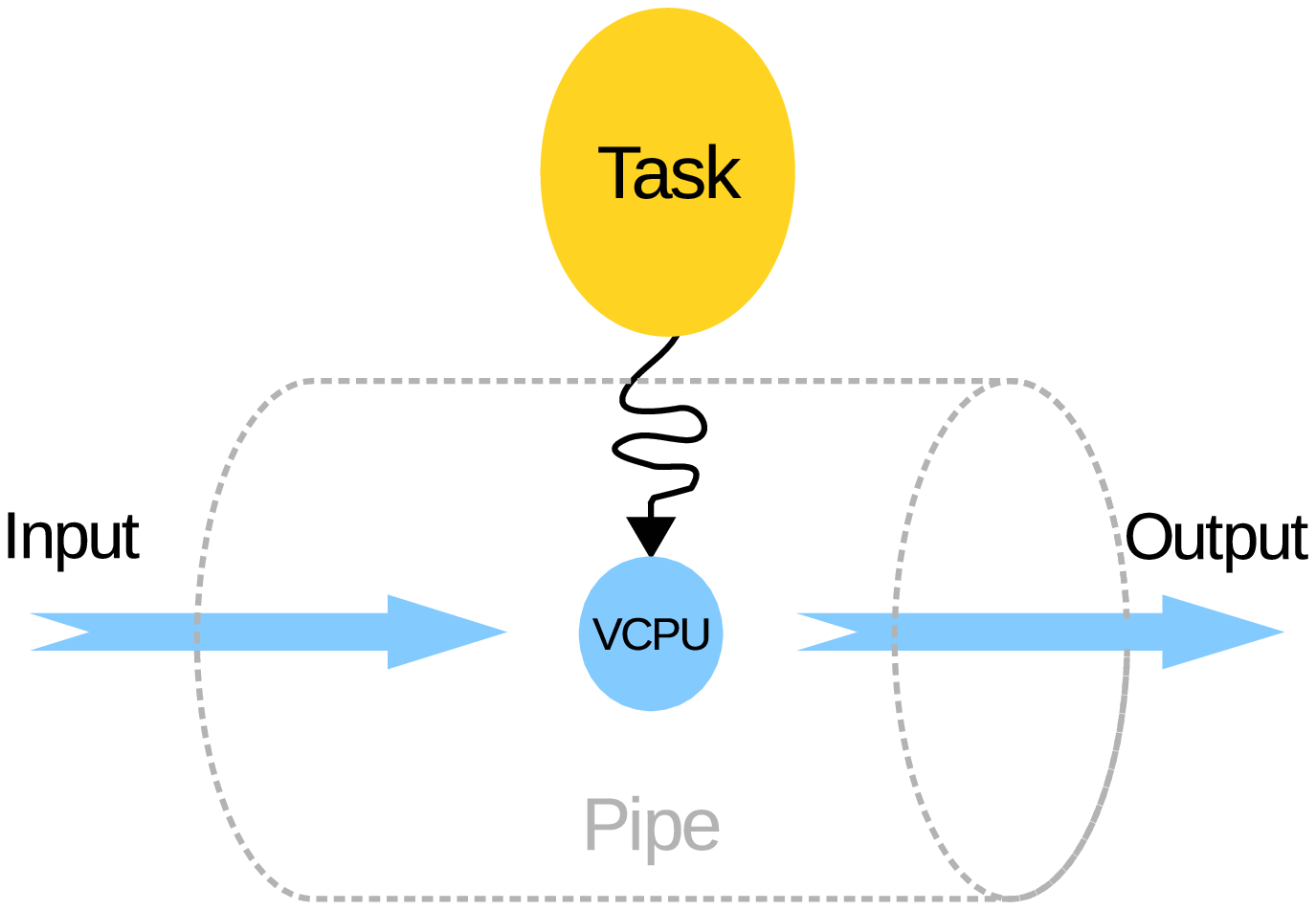}
	\caption{Illustration of a Pipe}
  \label{fig:pipe_concept}
\vspace{-0.15in}
\end{figure}

In Section~\ref{sect:latency_contr}, we identified the factors that influence
end-to-end communication delay. Among them, the absolute processing time and
the transfer data size are determined by the nature of the task in question.
To capture the rest of the timing characteristics, we develop a composable
pipe model, leveraging the scheduling approach described in
Section~\ref{sect:sched}. A task and pipe have a one-to-one relationship, as
illustrated in Figure~\ref{fig:pipe_concept}.

\setlength{\textfloatsep}{10pt}

\subsubsection{Terminology}
A pipe is composed of three elements:
\begin{itemize}
	\item \textbf{Pipe Terminal.} A pipe has one terminal, which is an
          encapsulation of the data processing power reserved for this pipe.
          A pipe terminal is represented by a VCPU and its timing
          characteristics are captured by the VCPU's budget and period,
          thereby guaranteeing at least $C$ units of execution time every $T$
          time units. Pipe terminals are associated with conventional tasks
          bound to Main VCPUs and kernel control paths (including interrupt
          handlers and device drivers) bound to I/O VCPUs, as described in
          Section~\ref{sect:sched}.
	\item \textbf{Pipe End.} A pipe has two ends, one for input and one
          for output. A pipe end is an interface to a communication channel,
          which is either an I/O bus or shared memory. Theoretically, the
          physical timing characteristics of a pipe end consist of
          transmission delay and propagation delay. As this work focuses on
          embedded systems where communicating parties are typically located
          within close proximity, we neglect propagation delay. The
          transmission delay is modeled by the bandwidth parameter, $W$, of
          the communication channel. We also use a parameter $\delta$ to
          denote the software overheads of a communication protocol. Though we
          are aware that $\delta$ depends on the data transfer size, the time
          difference is negligible, compared to the time of actual data
          transfer and processing.  Therefore, for the sake of simplicity,
          $\delta$ is a constant in our model.

\end{itemize}

 Note that in our definition of a single pipe there is only one terminal, not
 two at either end of the pipe. This differs from the idea of a POSIX pipe,
 which comprises at task at both sending and receiving ends. In our case, a
 pipe is represented by the single terminal that takes input and produces
 output.

\begin{figure}[!ht]
  \centering
  \includegraphics[scale=.6]{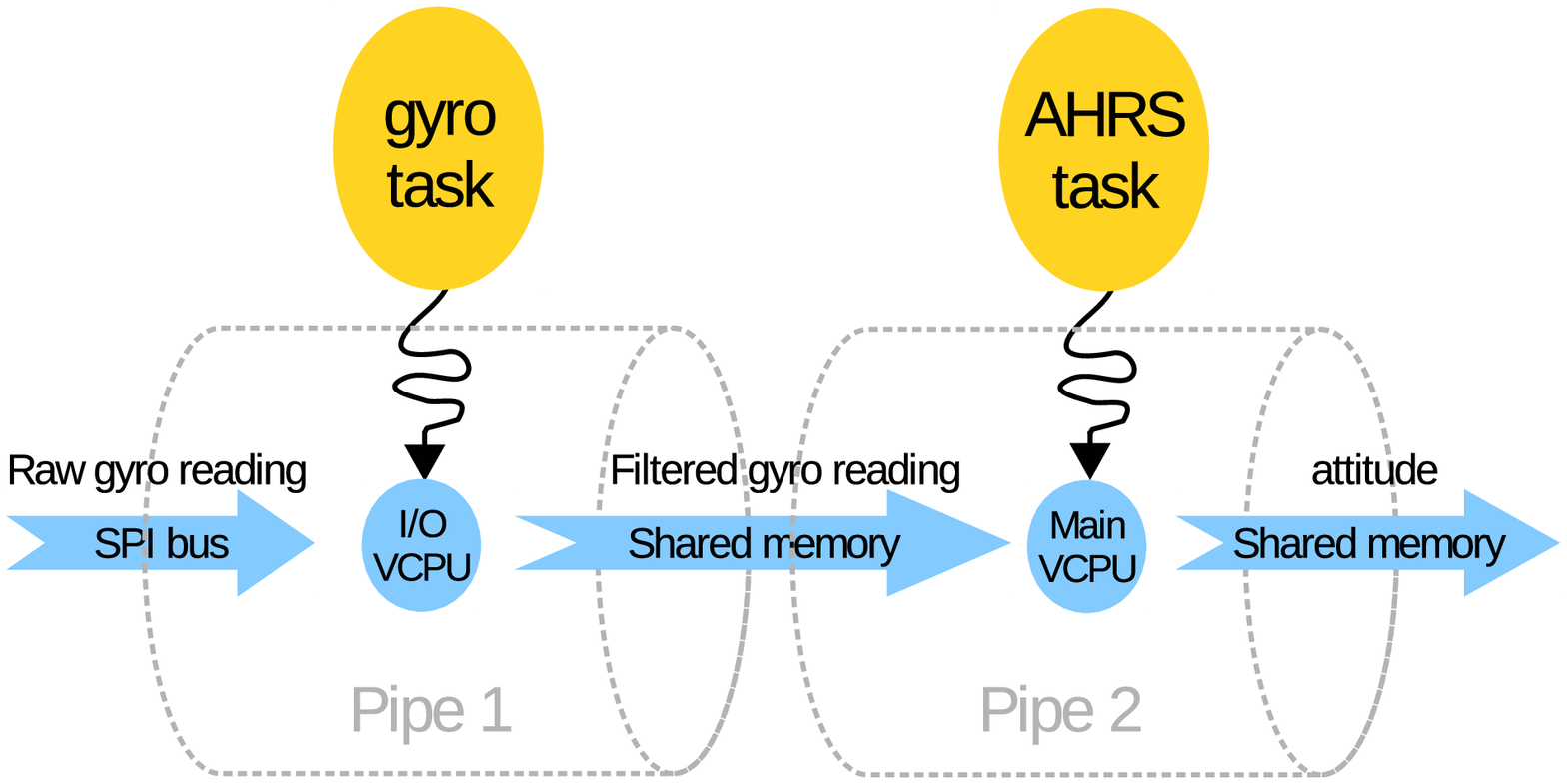}
  \caption{Illustration of Two Communicating Pipes}
  \label{fig:pipe_aero_example}
\vspace{-0.15in}
\end{figure}

 An example of two communicating pipes is shown in
 Figure~\ref{fig:pipe_aero_example}. This is representative of a communication
 path between a gyro task and attitude calculation in Cleanflight on the Aero
 board. The gyro task is mapped to Pipe 1, whose input end is over the SPI bus
 connected to the gyro sensor and output end is over a region of memory shared
 with Pipe 2. Pipe 1's terminal is an I/O VCPU because the gyro task is
 responsible for handling I/O interrupts generated from the SPI bus. On the
 contrary, the terminal of Pipe 2 is a Main VCPU as the AHRS task is
 CPU-intensive. The gyro task reads raw gyro readings from Pipe 1's input end,
 processes them, and writes filtered gyro readings to Pipe 1's output end.
 Similarly, the AHRS task reads the filtered gyro readings from Pipe 2's input
 end and produces attitude data for its output end over shared memory.

\subsubsection{Notation}
The timing characteristics of a pipe are denoted by the 3-tuple,
$\pi = ((W_{i}, \delta_{i}), (C, T), (W_{o}, \delta_{o}))$, where:
\begin{itemize}
\item $(W_{i}, \delta_{i})$ and $(W_{o}, \delta_{o})$  denote the bandwidth
  and software overheads of the input and output ends, respectively.
\item $(C, T)$ denotes the budget and period of the pipe terminal.
\end{itemize}

\noindent
A task $\tau$ is also represented as a 3-tuple, $\tau = (d_i, p, d_o)$, where:
\begin{itemize}
\item $d_{i}$ denotes the size of the raw data that is read in by $\tau$ in
  order to perform its job, and $d_{o}$ denotes the size of the processed data
  that is produced by $\tau$.
\item $p$ denotes the uninterrupted processing time it takes for $\tau$ to
  turn the raw data into the processed data.
\end{itemize}

\noindent
In addition, $\tau \mapsto \pi$ denotes the mapping between task $\tau$ and
pipe $\pi$.  A task $\tau = (d_i, p, d_o)$ is said to be mapped to a pipe $\pi
= ((W_{i}, \delta_{i}), (C, T), (W_{o}, \delta_{o}))$ when
\begin{itemize}
\item data of size $d_i$ is read from the input end with parameters
  $(W_{i}, \delta_{i})$, and data of size $d_o$ is written to the output
  end with parameters $(W_{o}, \delta_{o})$;
\item the pipe terminal with parameters $(C, T)$ is used for scheduling and
  accounting of the read and write operations, as well as the processing that
  takes time $p$.
\end{itemize}
For the composition of a chain of pipes, the operator $|$ connects a pipe's
output end to its succeeding pipe's input end. For example,
Figure~\ref{fig:pipe_aero_example} is represented as ${\tau_{gyro} \mapsto
  \pi_1}|{\tau_{AHRS} \mapsto \pi_2}$. The scheduling latency between two
pipes is denoted by $S_{\tau \mapsto \pi | \tau' \mapsto \pi'}$. Lastly, given
a task set $T=\{\tau_1, \tau_2, \cdots, \tau_n\}$ identity mapped to a pipe
set $\Pi = \{\pi_1, \pi_2, \cdots, \pi_n\}$, where pipes are connected to each
other in ascending order of subscript, $E_{\tau_1 \mapsto \pi_1 | \tau_2
  \mapsto \pi_2 | \cdots | \tau_n \mapsto \pi_n}$ denotes the end-to-end
reaction time of the pipe chain, and $F_{\tau_1 \mapsto \pi_1 | \tau_2 \mapsto
  \pi_2 | \cdots | \tau_n \mapsto \pi_n}$ denotes the end-to-end freshness
time.



\subsection{Reachability} 
\label{sect:reachability}
Before mathematically analyzing end-to-end time, we introduce the concept of
\verb+reachability+, inspired by the \textit{data-path reachability
  conditions} proposed by Feiertag et al~\cite{Feiertag2008}.  The necessity
of introducing reachability is due to a subtle difference between our
register-based asynchronous communication model and the traditional FIFO-based
synchronous communication. In the latter, data is guaranteed to be transferred
without loss or repetition. This way, end-to-end time is derived from the time
interval between the arrival of a data input and the departure of its {\em
  corresponding} data output. Unfortunately, this might result in an
infinitely large end-to-end time in the case of register-based asynchronous
communication where not every input leads to an output.  Instead, unprocessed
input data might be discarded (overwritten) when newer input data is
available, as explained in Section~\ref{sect:async_comm}.

An infinitely large end-to-end time, while mathematically correct, lacks
practical use. Therefore, the following timing analysis ignores all input data
that fails to ``reach'' the exit of the pipe chain it enters. Instead, only
those data inputs that result in data outputs from the pipe chain are
considered. We define this latter class of inputs as being \verb+reachable+.


\subsection{Timing Analysis}
As alluded to above, the execution of a task is divided into three
stages, involving (1) reading, (2) processing, and (3) writing data.
To simplify the timing analysis, we assume that tasks are able to finish the
read and write stages within one period of the pipe terminal, to which the
task is mapped.  This is not unrealistic for applications such as a flight
controller, because: 1) data to be transferred is usually small, and 2) all
three stages are typically able to finish within one period. However, to
maintain generality, we do not impose any restriction on the length of the
processing stage.


\subsubsection{Worst Case End-to-end Time of a Single Pipe}
\label{sect:single_pipe}
First, we consider the case where there is a single pipe. Two key observations
for this case are: 1) the absence of scheduling latency due to the lack of a
succeeding pipe, and 2) the equivalence of the two end-to-end time semantics
(reaction and freshness time) due to the lack of a preceding pipe. We
therefore use $L_{\tau \mapsto \pi}$ to unify the notation of $E_{\tau \mapsto
  \pi}$ and $F_{\tau \mapsto \pi}$.

Given task $\tau = (d_i, p, d_o)$ mapped to pipe $\pi = ((W_{i}, \delta_{i}),
(C, T), (W_{o}, \delta_{o}))$, the worst case end-to-end time is essentially
the execution time of the three stages of $\tau$ on $\pi$. Due to the timing
property of $\pi$'s pipe terminal, $\tau$ is guaranteed $C$ units of execution
time within any window of $T$ time units. Hence, the worst-case latency
$L_{\tau \mapsto \pi}^{wc}$ is bounded by the following:
\begin{align}\label{equ:worst_case}
	L_{\tau \mapsto \pi}^{wc} = {\left\lfloor\frac{\Delta_{in} + p +
            \Delta_{out}}{C}\right\rfloor} \cdot T + (\Delta_{in} + p +
        \Delta_{out}) \mod C
\end{align}
where $\Delta_{in} = \frac{d_i}{W_i} + \delta_i$ and $\Delta_{out} =
\frac{d_o}{W_o} + \delta_o$.

\subsubsection{Worst Case End-to-end Reaction Time of a Pipe Chain}
\label{sect:reaction}
In this section, we extend the timing analysis of a single pipe to a pipe
chain.  For the sake of simplicity, we start with a chain of length two. We
show in Section~\ref{sect:composability} that the mathematical framework is
applicable to arbitrarily long pipe chains.  To distinguish the tasks mapped
to the two pipes, we name the preceding task \verb+producer+ and the
succeeding \verb+consumer+.  The producer is denoted by $\tau_p=(d_{i}^{p},
p^{p}, d)$ and its pipe is denoted by $\pi_p = ((W_{i}^{p}, \delta_{i}^{p}),
(C^{p}, T^{p}), (W_{o}^{p}, \delta_{o}^{p}))$.  Similarly, the consumer task
and pipe are denoted by $\tau_c=(d,p^{c},d_{o}^{c})$ and $\pi_c = ((W_{i}^{c},
\delta_{i}^{c}), (C^{c}, T^{c}), (W_{o}^{c}, \delta_{o}^{c}))$.  Following the
definition of the end-to-end reaction time, $E_{{\tau_p \mapsto \pi_p}|{\tau_c
    \mapsto \pi_c}}$, in Section~\ref{sect:semantic}, we investigate the time
interval between a specific instance of input data, denoted by $D_i$, being
read by $\tau_p$, and its \textbf{first} corresponding output, denoted by
$D_o$, being written by $\tau_c$.

It is of vital importance to recognize that end-to-end time of a pipe chain is
not simply the sum of the end-to-end time of each single pipe in the chain. We
also need to account for the \verb+scheduling latency+ resulting from each
appended pipe.  As described in Section~\ref{sect:latency_contr}, the
scheduling latency depends on the order of execution of tasks. We, therefore,
perform the timing analysis under two complementary cases: \textbf{Case 1 -}
$\tau_c$ has shorter period and thus higher priority than $\tau_p$;
\textbf{Case 2 -} $\tau_p$ has shorter period and thus higher priority than
$\tau_c$, according to rate-monotonic ordering.

\paragraph{Calculating the End-to-end Reaction Time}
\label{sect:calculation_reaction}

\begin{figure}[!ht]
  \centering \includegraphics[scale=0.5]{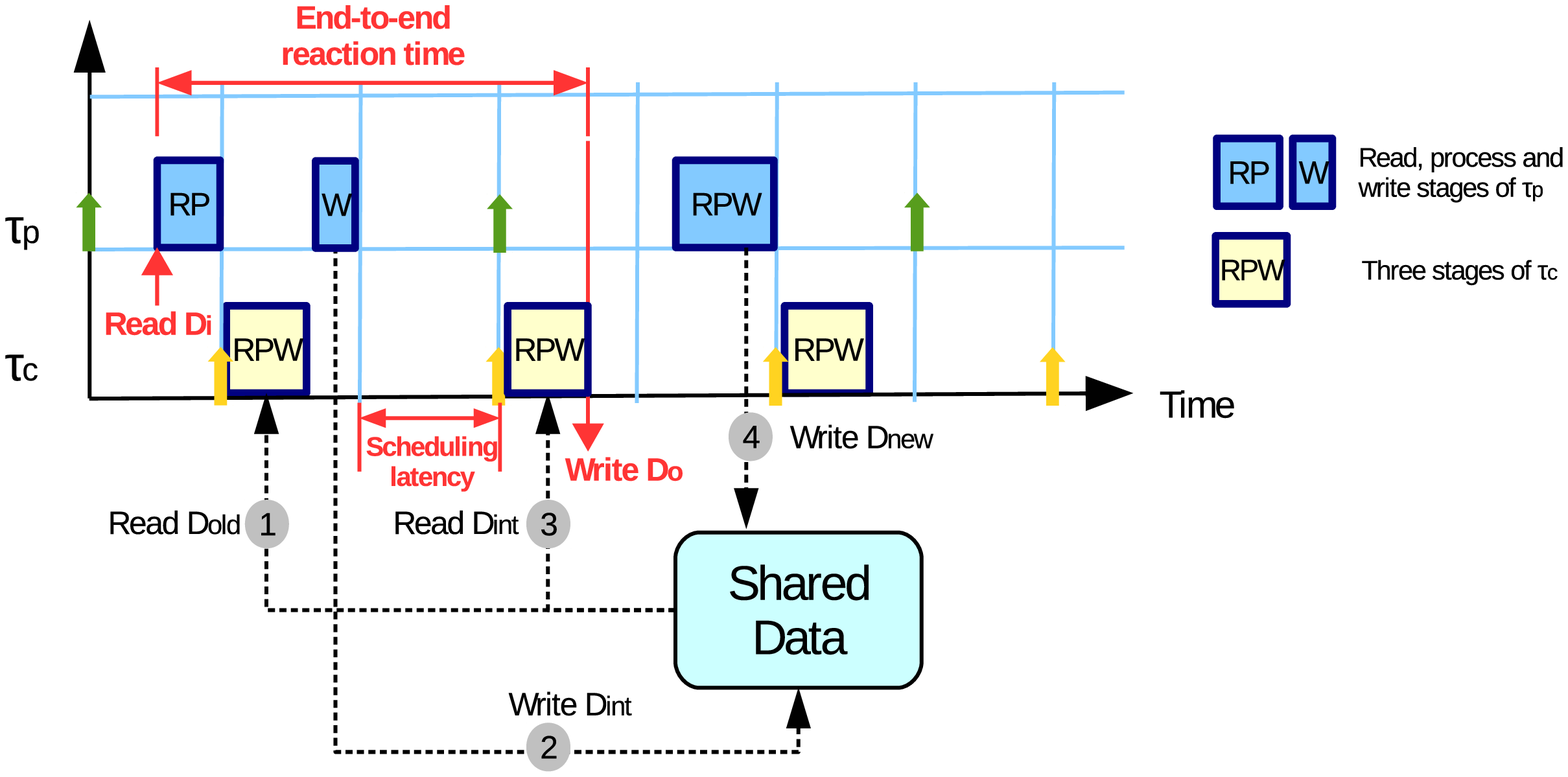}
  \caption{End-to-end Reaction Time in Case 1}
  \label{fig:e2e_reaction0}
\vspace{-0.15in}
\end{figure}

\textbf{Case 1.}  The key to making use of $L_{\tau_p \mapsto \pi_p}^{wc}$ and
$L_{\tau_c \mapsto \pi_c}^{wc}$ in the timing analysis of $E_{{\tau_p \mapsto
    \pi_p}|{\tau_c \mapsto \pi_c}}^{wc}$, is to find the worst case scheduling
latency, $S_{{\tau_p \mapsto \pi_p}|{\tau_c \mapsto \pi_c}}^{wc}$.  As
illustrated in Figure~\ref{fig:e2e_reaction0}, the worst case scheduling
latency occurs when $\tau_c$ preempts $\tau_p$ (Step \circled{1}) immediately
before $\tau_p$ produces the intermediate output $D_{int}$ corresponding to
$D_i$.  After preemption, $\tau_c$ uses up $\pi_c$'s budget and gives the CPU
back to $\tau_p$. Upon being resumed, $\tau_p$ immediately produces $D_{int}$
(Step \circled{2}). For $\tau_c$ to become runnable again to read $D_{int}$ in
Step \circled{3}, it has to wait for its budget replenishment. The waiting
time is exactly the worst case scheduling latency:
\begin{align}
  S_{{\tau_p \mapsto \pi_p}|{\tau_c \mapsto \pi_c}}^{wc} = T^{c} - C^{c}
  - (\frac{d}{W_{o}^{p}} + \delta_{o}^{p})
\end{align}

After replenishment, $\tau_c$ reads in $D_{int}$, processes it and eventually
writes out $D_o$. As $E_{{\tau_p \mapsto \pi_p}|{\tau_c \mapsto \pi_c}}$ is
defined to be the time interval between the arrival of $D_i$ and the departure
of $D_o$, the worst case of $E_{{\tau_p \mapsto \pi_p}|{\tau_c \mapsto
    \pi_c}}$ is as follows:
\begin{align} \label{equ:e2e_reaction0}
  \begin{split}
    E_{{\tau_p \mapsto \pi_p}|{\tau_c \mapsto \pi_c}}^{wc} & =
    L_{\tau_p \mapsto \pi_p}^{wc} + S_{{\tau_p \mapsto
        \pi_p}|{\tau_c \mapsto \pi_c}}^{wc} + L_{\tau_c \mapsto
      \pi_c}^{wc}\\ & = L_{\tau_p \mapsto \pi_p}^{wc} + L_{\tau_c
      \mapsto \pi_c}^{wc} + T^{c} - C^{c} - (\frac{d}{W_{o}^{p}} +
    \delta_{o}^{p})
  \end{split}
\end{align}
Note that if $\tau_c$ runs out of budget before writing $D_{o}$, $\tau_p$
may overwrite $D_{int}$ in the pipe with new data (Step \circled{4}).
However, the {\em implicit communication} property guarantees that $\tau_c$
only works on its local copy of the shared data, which is $D_{int}$ until
$\tau_c$ initiates another read.

\textbf{Case 2.} The situation is more complicated when $\tau_p$ has higher
priority than $\tau_c$. The worst case scenario in Case 1 does not hold in
Case 2 primarily because $\tau_p$ might overwrite $D_{int}$ before $\tau_c$
has a budget replenishment. This is impossible in Case 1 because $\tau_p$ has
a larger period than $\tau_c$, which is guaranteed to have
its budget replenished before $\tau_p$ is able to initiate another write. In
other words, in Figure~\ref{fig:e2e_reaction0}, Step \circled{3} is guaranteed
to happen before Step \circled{4}.

\begin{figure}[t!]
  \centering \includegraphics[scale=0.5]{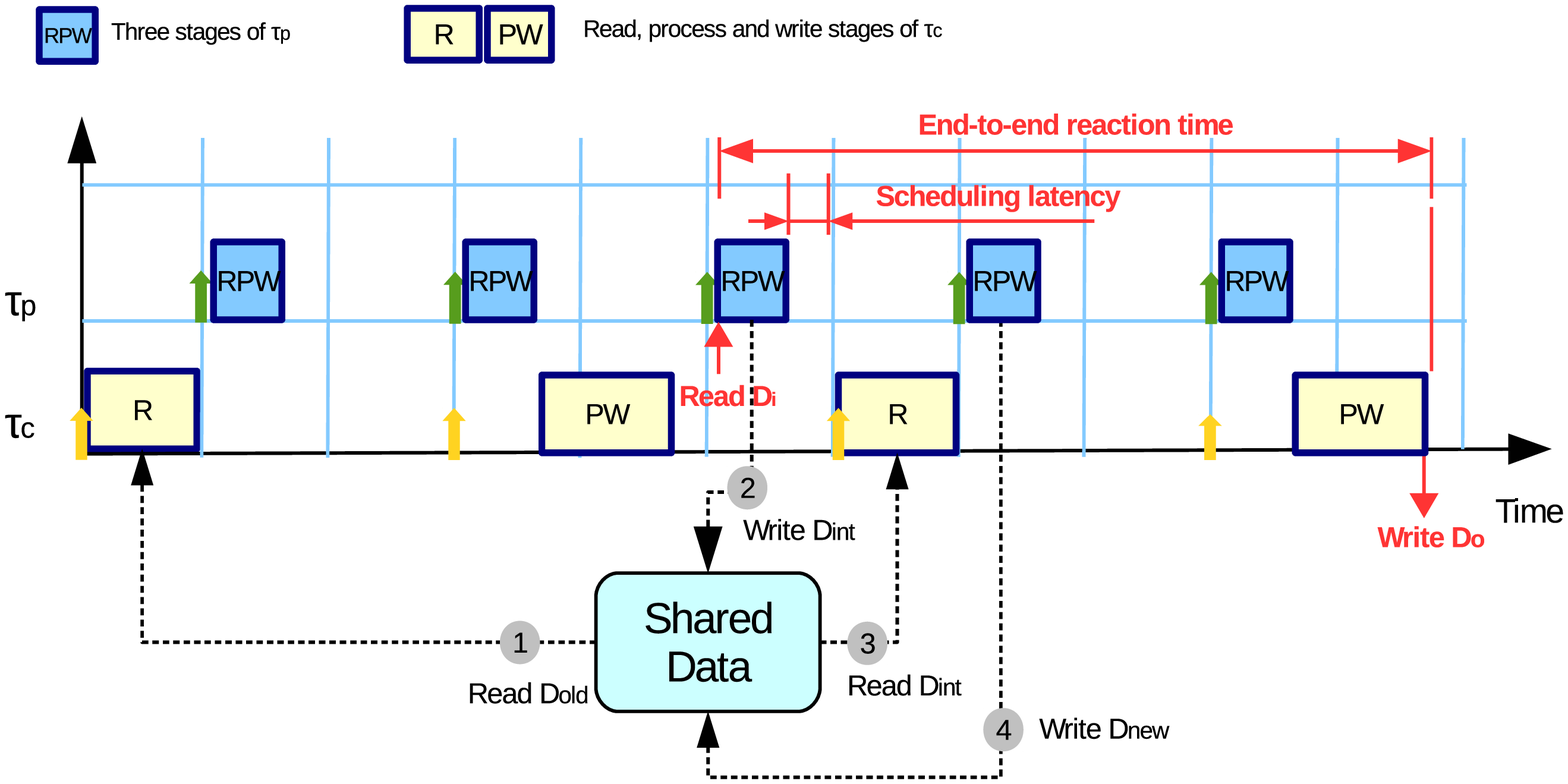}
  \caption{End-to-end Reaction Time in Case 2}
  \label{fig:e2e_reaction1}
  \vspace{-0.15in}
\end{figure}

The data-overwrite problem in Case 2 is the reason for introducing
reachability in Section~\ref{sect:reachability}. To find the worst case
end-to-end reaction time in this case, we have to find the scenario that not
only leads to the worst case scheduling latency, but also originates from a
\verb+reachable+ input.  Figure~\ref{fig:e2e_reaction1} illustrates a scenario
that meets these requirements.  In the figure, $\tau_p$ preempts $\tau_c$
immediately after $\tau_c$ finishes reading $\tau_p$'s intermediate output
(Step \circled{3}), $D_{int}$, corresponding to $D_i$.  It follows that the
longest possible waiting time, between $D_{int}$ becoming available (Step
\circled{2}) and $\tau_c$ reading the data (Step \circled{3}), is the period
of $\tau_p$ minus both its budget and the execution time of the read stage of
$\tau_c$.  This waiting time is exactly the worst case scheduling latency:
\begin{align} \label{equ:sched_latency1}
  S_{{\tau_p \mapsto \pi_p}|{\tau_c \mapsto \pi_c}}^{wc} =
  T^{p} - C^{p} - (\frac{d}{W_{i}^{c}} + \delta_{i}^{c})
\end{align}

Between reading $D_{int}$ and writing $D_o$, $\tau_c$ might experience
more than one preemption from $\tau_p$, which repeatedly overwrites the shared
data. This will not, however, affect $\tau_c$'s processing on $D_{int}$ either
spatially or temporally, thanks to the VCPU model and the implicit
communication semantic.  Therefore, similar to Case 1, the worst case
end-to-end reaction time is again the sum of Equation~\ref{equ:worst_case} of
each pipe and Equation~\ref{equ:sched_latency1}:
\begin{align} \label{equ:e2e_reaction1}
  \begin{split}
    E_{{\tau_p \mapsto \pi_p}|{\tau_c \mapsto \pi_c}}^{wc} & =
    L_{\tau_p \mapsto \pi_p}^{wc} + S_{{\tau_p \mapsto
        \pi_p}|{\tau_c \mapsto \pi_c}}^{wc} + L_{\tau_c \mapsto
      \pi_c}^{wc}\\ & = L_{\tau_p \mapsto \pi_p}^{wc} + L_{\tau_c
      \mapsto \pi_c}^{wc} + T^{p} - C^{p} - (\frac{d}{W_{i}^{c}} +
    \delta_{i}^{c})
  \end{split}
\end{align}

Since the output end of $\tau_p$ and the input end of $\tau_c$ share the same
communication channel, it is reasonable to assume that $W_{o}^{p} = W_{i}^{c}$
and $\delta_{o}^{p} = \delta_{i}^{c}$. With that, we proceed to unify the
worst case end-to-end reaction time using one conditional equation:
\begin{align} \label{equ:reaction}
  E_{{\tau_p \mapsto \pi_p}|{\tau_c \mapsto \pi_c}}^{wc} =
  \begin{cases}
    T^{c} - C^{c} - (\frac{d}{W} + \delta) + L_{\tau_p \mapsto \pi_p}^{wc} +  L_{\tau_c \mapsto \pi_c}^{wc}, & \text{if } T^{c} < T^{p}\\
    T^{p} - C^{p} - (\frac{d}{W} + \delta) + L_{\tau_p \mapsto \pi_p}^{wc} +  L_{\tau_c \mapsto \pi_c}^{wc}, & \text{otherwise}
  \end{cases}
\end{align}
where $W = W_{o}^{p} = W_{i}^{c}$ and $\delta = \delta_{o}^{p} = \delta_{i}^{c}$

\paragraph{Special Cases}
\label{sect:simplification}
Real-time systems are often profiled offline to obtain worst case execution
times of their tasks. In our case, this would enable CPU resources for pipe
terminals to be provisioned so that each task completes one iteration of all
three stages (read, process, write) in one budget allocation and, hence,
period. This implies that $\Delta_{in} + p + \Delta_{out} + \epsilon= C$ in
Equation~\ref{equ:worst_case}, where $\epsilon$ is an arbitrarily small
positive number to account for surplus budget after completing all task
stages.  With that, it is possible to simplify the worst case end-to-end
reaction time derived in Section~\ref{sect:calculation_reaction}. First,
Equation~\ref{equ:worst_case} is simplified as follows:
\begin{align}\label{equ:simple_worst_case}
  \begin{split}
    L_{\tau \mapsto \pi}^{wc} & =
    \lfloor\frac{C-\epsilon}{C}\rfloor \cdot T + [(C - \epsilon)
      \mod C]\\ & = 0 \cdot T + (C - \epsilon) \approx C
  \end{split}
\end{align}
Using Equation~\ref{equ:simple_worst_case}, Equation~\ref{equ:e2e_reaction1}
reduces to:
\begin{align} \label{equ:simple_e2e_reaction1}
  \begin{split}
    E_{{\tau_p \mapsto \pi_p}|{\tau_c \mapsto \pi_c}}^{wc} & =
    T^{p} - C^{p} - \Delta_{in}^{c} + L_{\tau_p \mapsto
      \pi_p}^{wc} + L_{\tau_c \mapsto \pi_c}^{wc}\\ & = T^{p} -
    C^{p} - \Delta_{in}^{c} + C^{p} + C^{c} \\ & = T^{p} + C^{c} -
    \Delta_{in}^{c}
  \end{split}
\end{align}
The same simplification applied to Equation~\ref{equ:e2e_reaction0} of
Case 1 reduces Equation~\ref{equ:reaction} to:
\begin{align}\label{equ:e2e_simple_reaction}
  E_{{\tau_p \mapsto \pi_p}|{\tau_c \mapsto \pi_c}}^{wc} =
  \begin{cases}
    T^{c} + C^{p} -\Delta,		 & \text{if } T^{c} < T^{p}\\
    T^{p} + C^{c} - \Delta,		 & \text{otherwise}
  \end{cases}
\end{align}
where $\Delta = \frac{d}{W} + \delta$.

If we further assume that $\pi_p$ and $\pi_c$ communicate data of small size
over shared memory, it is possible to discard communication overheads, such
that $\Delta = 0$.  With that, Equation~\ref{equ:e2e_simple_reaction}
simplifies to:
\begin{align}\label{equ:e2e_simple_simple_reaction}
  E_{{\tau_p \mapsto \pi_p}|{\tau_c \mapsto \pi_c}}^{wc} =
  \begin{cases}
    T^{c} + C^{p},		 & \text{if } T^{c} < T^{p}\\
    T^{p} + C^{c},		 & \text{otherwise}
  \end{cases}
\end{align}
Finally, notice that by appending $\tau_c \mapsto \pi_c$ to $\tau_p \mapsto
\pi_p$, the worst case end-to-end reaction time is increased by the
following:
\begin{align}\label{equ:e2e_simple_reaction_inc}
  \begin{split}
    \uparrow{E^{wc}} = E_{{\tau_p \mapsto \pi_p}|{\tau_c \mapsto
        \pi_c}}^{wc} - E_{\tau_p \mapsto \pi_p}^{wc} &= E_{{\tau_p
        \mapsto \pi_p}|{\tau_c \mapsto \pi_c}}^{wc} -
    C^{p}\\ &= \begin{cases} T^{c}, & \text{if } T^{c} <
      T^{p}\\ T^{p} - C^{p} + C^{c}, & \text{otherwise}
    \end{cases}
  \end{split}
\end{align}

\subsubsection{Worst Case End-to-end Freshness Time of a Pipe Chain}
\label{sect:freshness}
Techniques similar to those in Section~\ref{sect:reaction} will be used to
analyze end-to-end freshness time. To avoid repetition, we abbreviate the
end-to-end freshness timing analysis by only focusing on the special cases
described in Section~\ref{sect:simplification}.

Recall that freshness time is defined to be the interval between the
arrival of an input and the departure of its {\em last corresponding}
output. Therefore, we investigate the interval between a specific
instance of input data, $D_i$, being read by $\tau_p$ and its
{\em last} corresponding output, $D_o$, being written by
$\tau_c$.

\begin{figure}[!ht]
  \centering \includegraphics[scale=0.5]{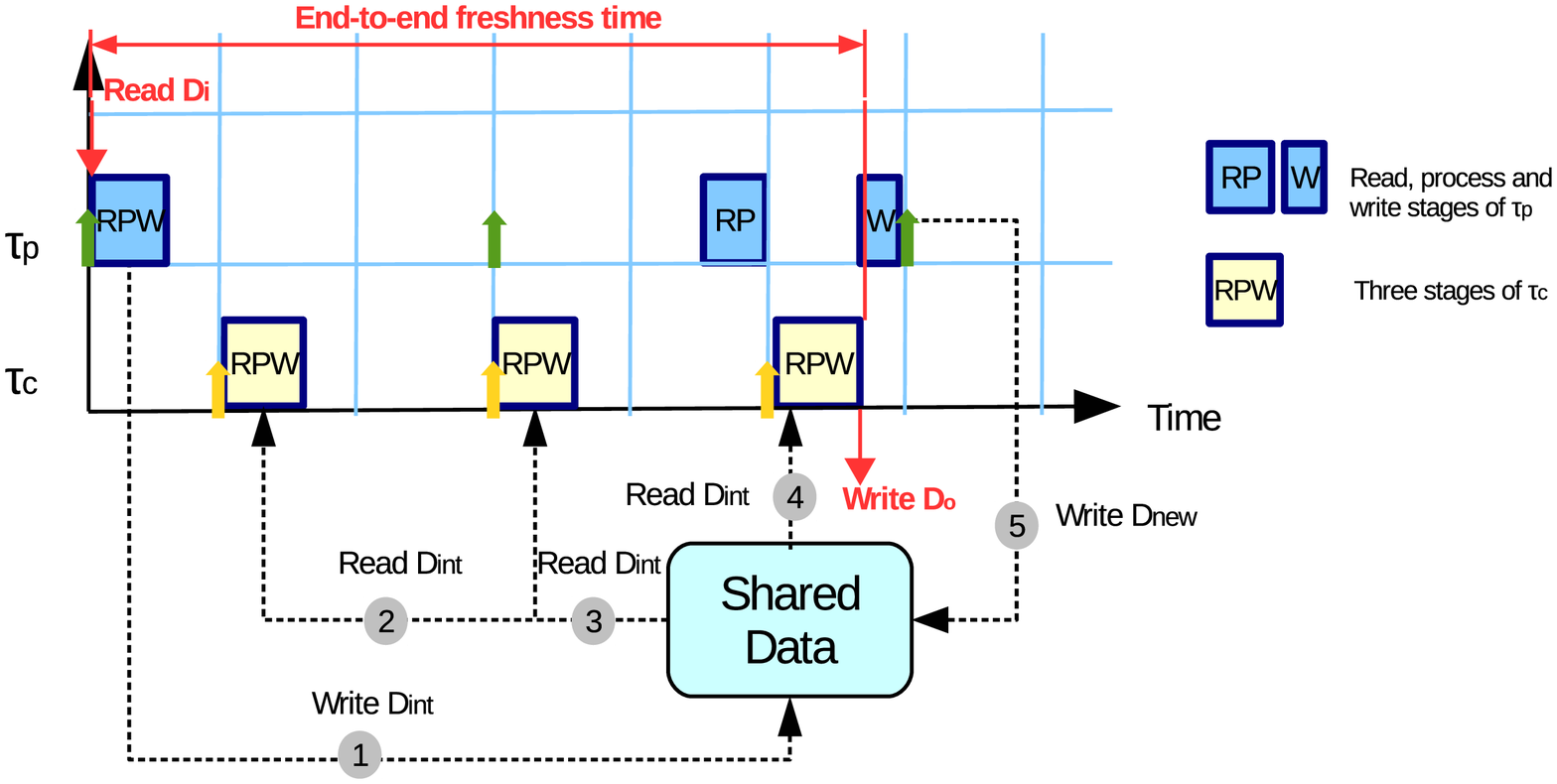}
  \caption{End-to-end Freshness Time in Case 1}
  \label{fig:e2e_fresh0}
  \vspace{-0.15in}
\end{figure}

\textbf{Case 1.} As illustrated in Figure~\ref{fig:e2e_fresh0}, $D_i$ is read
by the first instance of $\tau_p$ at time 0 and the intermediate output,
$D_{int}$, is written to the shared data (Step \circled{1}).  After that,
$\tau_c$ produces three outputs corresponding to $D_{int}$ (Steps \circled{2},
\circled{3} and \circled{4}), or to $D_i$ indirectly.  The last output, $D_o$,
is the one preceding $\tau_p$'s write of new data, $D_{new}$ (Step
\circled{5}).  Thus, the worst case end-to-end freshness time, $F_{{\tau_p
    \mapsto \pi_p}|{\tau_c \mapsto \pi_c}}^{wc}$, occurs when: 1) the two
consecutive writes (Steps \circled{1} and \circled{5}) from $\tau_p$ have the
longest possible time interval between them, and 2) the write of $D_o$ happens
as late as possible. The latest time to write $D_o$ is immediately before the
second write of $\tau_p$, which is preempted by higher priority $\tau_c$.

From Figure~\ref{fig:e2e_fresh0} that the worst case end-to-end freshness time
is:
\begin{align}\label{equ:e2e_fresh0}
  F_{{\tau_p \mapsto \pi_p}|{\tau_c \mapsto \pi_c}}^{wc} = 2 \cdot T^{p}
  - \Delta_{out}^{p}
\end{align}
Again, when communicating over shared memory, Equation~\ref{equ:e2e_fresh0}
can be further simplified to:
\begin{align}\label{equ:e2e_simple_fresh0}
  F_{{\tau_p \mapsto \pi_p}|{\tau_c \mapsto \pi_c}}^{wc} = 2 \cdot T^{p}
\end{align}

\textbf{Case 2.} When $\tau_p$ has a smaller period than $\tau_c$, it is
impossible for $\tau_c$ to read the same intermediate output of $\tau_p$
twice. In Figure~\ref{fig:e2e_fresh0}, Step~\circled{5} is guaranteed to
happen before \circled{3}. Thus, the worst case freshness time is essentially
the worst case reaction time, shown in Equation~\ref{equ:e2e_reaction1}.

In summary, the worst case end-to-end freshness latency of two communicating
pipes is represented in the following conditional equation:
\begin{align}\label{equ:e2e_simple_fresh}
  F_{{\tau_p \mapsto \pi_p}|{\tau_c \mapsto \pi_c}} =
  \begin{cases}
    2 \cdot T^{p}, & \text{if } T^{c} < T^{p}\\ T^{p} +
    C^{c}, & \text{otherwise}
  \end{cases}
\end{align}

\subsection{Composability}
\label{sect:composability}
The timing analysis for two pipes in Section~\ref{sect:reaction} extends to
pipe chains of arbitrary length.  Every time an extra task $\tau_{new}$
(mapped to $\pi_{new}$) is appended to the tail end of a chain ($\tau_{tail}
\mapsto \pi_{tail}$), the worst case end-to-end reaction time increases by the
worst case end-to-end time of the newly appended pipe, plus the scheduling
latency between the new pipe and the tail pipe.  The actual value of the
increase, depending on the relative priority of the new pipe and the tail
pipe, is shown in Equation~\ref{equ:e2e_simple_reaction_inc}. Similarly, the
added end-to-end freshness time can be derived from
Equation~\ref{equ:e2e_simple_fresh}.

Composability is a crucial property of our pipe model, since it
significantly eases the end-to-end time calculation for any given
pipeline. This provides the basis for a design framework that derives task
periods from given end-to-end timing constraints. This is detailed in the
following section.

\section{End-to-end Design}
\label{sect:e2e_design}
There are significant challenges to porting a flight control firmware such as
Cleanflight to run as a multithreaded application on a real-time operating
system. One of the major issues is how to determine the period of each thread
so that the application is able to meet its end-to-end timing constraints.  A
naive approach would be to start by choosing a tentative set of periods and
use the timing analysis method in Section~\ref{sect:e2e_analysis} to validate
the timing correctness. Upon failure, the periods are heuristically adjusted
and the validation step is repeated until end-to-end timing guarantees are
met. This approach, however, is potentially time-consuming and labor-intensive
when the number of tasks or end-to-end constraints increase.

Inspired by Gerber \textit{et al} \cite{Gerber1995}, we derive task periods
from end-to-end timing constraints, by combining the timing analysis of the
pipe model with linear optimization techniques.  In this section, we
generalize our method for use with a broader spectrum of cyber-physical
control applications.

\subsection{Problem Definition}
\begin{figure}[t!]
	\begin{floatrow} \ffigbox{%
		 \includegraphics[scale=0.3]{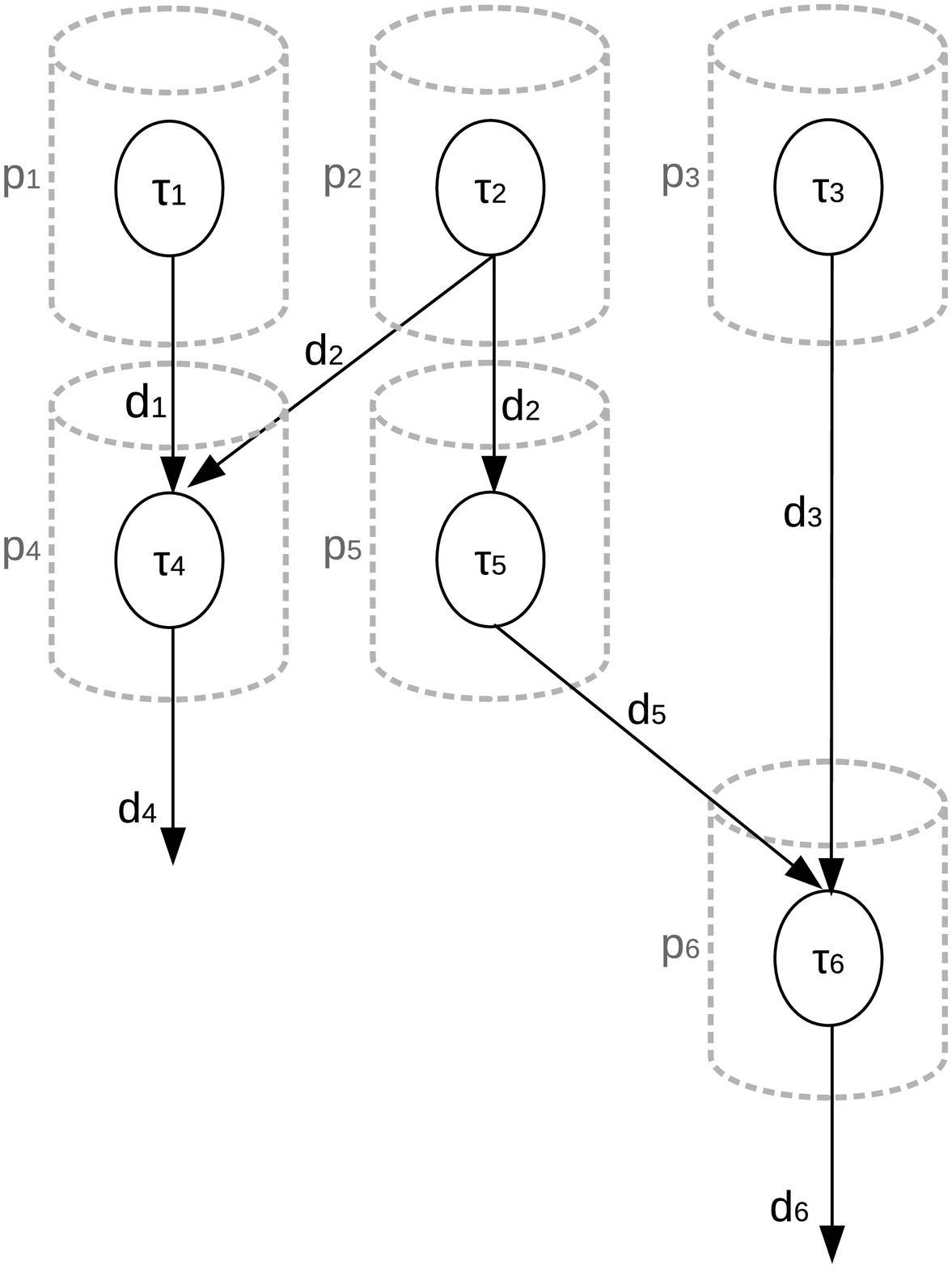}
		}{%
			\caption{Application Task Graph}%
		  \label{fig:task_graph}
		}
		\capbtabbox{%
			\small
			\begin{tabular}{c|c} \hline
				& \multirow{6}{*}{\texttt{Reaction}} \\
				$E_{{\tau_1 \mapsto \pi_1}|{\tau_4 \mapsto \pi_4}} \leq 10$,\\
				$E_{{\tau_2 \mapsto \pi_2}|{\tau_4 \mapsto \pi_4}} \leq 15$,\\
				$E_{{\tau_2 \mapsto \pi_2}|{\tau_5 \mapsto \pi_5}|{\tau_6 \mapsto \pi_6}} \leq 25$,\\
				$E_{{\tau_3 \mapsto \pi_3}|{\tau_6 \mapsto \pi_6}} \leq 15$; \\
				\\ \hline
				& \multirow{6}{*}{\texttt{Freshness}} \\
				$F_{{\tau_1 \mapsto \pi_1}|{\tau_4 \mapsto \pi_4}} \leq 20$,\\
				$F_{{\tau_2 \mapsto \pi_2}|{\tau_4 \mapsto \pi_4}} \leq 30$,\\
				$F_{{\tau_2 \mapsto \pi_2}|{\tau_5 \mapsto \pi_5}|{\tau_6 \mapsto \pi_6}} \leq 50$,\\
				$F_{{\tau_3 \mapsto \pi_3}|{\tau_6 \mapsto \pi_6}} \leq 20$; \\
				\\ \hline
				& \multirow{3}{*}{\texttt{Schedulability}} \\
				$\sum_{j=1}^{6}\frac{C^{j}}{T^{j}} \leq n(\sqrt[6]{2}-1)$ \\ 
				\\ \hline
				& \multirow{7}{*}{\texttt{Execution Times}} \\
				$\forall j \in \{1,2,\cdots, 6\}$, \\
				$d_i^{j} = d_o^{j} = 3$, \\
				$W_i^{j} = W_o^{j} = 20$,\\
				$\delta_i^{j} = \delta_o^{j} = 0.1$,\\
				$p^{j}=0.5$;\\
				\\ \hline
			\end{tabular}
		}{%
			\caption{Application Timing Characteristics}%
		  \label{tab:example_timing}
		}
	\end{floatrow}
\end{figure}

To precisely define the problem, first consider a set of tasks $T
= \{\tau_1, \tau_2, \cdots, \tau_n\}$ and a set of pipes $\Pi
= \{\pi_1, \pi_2, \cdots, \pi_n\}$, where $\tau_j = (d_i^{j}, p^{j},
d_o^{j})$ and $\pi_j = ((W_{i}^{j}, \delta_{i}^{j}), (C^{j},\linebreak
T^{j}), (W_{o}^{j}, \delta_{o}^{j}))$. We additionally require the
following information:
\begin{itemize}
	\item the mapping between $T$ and $\Pi$. For ease of notation, we assume 
		tasks map to the pipe with the same subscript, hence $\forall j \in
		\{1,2,\cdots,n\}, \tau_j \mapsto \pi_j$;
	\item the topology of $\Pi$ (an example is shown in Figure~\ref{fig:task_graph});
	\item $\forall j \in \{1, 2, \cdots, n\}$, the value of $d_i^{j}$, $p^{j}$ and
	$d_o^{j}$;
	\item $\forall j \in \{1, 2, \cdots, n\}$, the value of $W_{i}^{j}$,
		$\delta_{i}^{j}$, $W_{o}^{j}$ and $\delta_{o}^{j}$;
	\item the end-to-end timing constraints, namely the value of 
	$E_{{\tau_i \mapsto \pi_i}|{\tau_j \mapsto \pi_j}|\cdots|{\tau_k \mapsto \pi_k}}$
		and/or 
	$F_{{\tau_p \mapsto \pi_p}|{\tau_q \mapsto \pi_q}|\cdots|{\tau_r \mapsto \pi_r}}$
		where $i, j, k, p, q, r \in \{1,2,\cdots,n\}$.
\end{itemize}

\noindent
The aim is to find a feasible set of $\{(C^{j}, T^{j})\}$ pairs for $j \in
\{1,2,\cdots,n\}$ that:
\begin{enumerate}
	\item meets all the specified end-to-end timing constraints,
        \item passes the task schedulability test, and
        \item ideally but not necessarily minimizes CPU utilization. 
        A task should not be run faster than it needs to be, so that 
        resources are made available for additional system objectives.
\end{enumerate}

\subsection{Solving the Constraints}

Our solution is carried out in a three-step process. To make it easier to
understand, we use a concrete example with actual numbers to elaborate the
process.  Consider the pipe topology graph shown in
Figure~\ref{fig:task_graph}, in which there are six tasks mapped to six
pipes. Tasks 1, 2 and 3 read inputs from sensors, Tasks 4 and 6 write their
outputs to actuators, and Task 5 is an intermediary responsible for
complicated processing such as PID control or sensor data fusion. The timing
characteristics of the tasks and pipes are shown in
Table~\ref{tab:example_timing}.  Note that the execution times are assumed to
be identical for all tasks. In practice this would not necessarily be the case
but it does not affect the generality of the approach.

In Step 1, we use the given $d_i$, $d_o$, $p$, $W_i$, $\delta_i$, $W_o$ and
$\delta_o$ to compute the budget of each pipe terminal.  The budget is set to
a value that ensures the three stages (\textit{i.e.}, read, process and write)
finish in one period.  To compute $C^{1}$, for example, we aggregate the
times for $\tau_1$ to read, process and write data. Thus $C^{1} =
\frac{d_i^{1}}{W_i^{1}} + \delta_i^{1} + p^{1} +
\frac{d_o^{1}}{W_o^{1}} + \delta_o^{1}$. All budgets are
computed in a similar way. \annotate{When the input to a pipe terminal comes
from multiple sources the value $d_i$ is aggregated from all input
channels. For example, $\tau_4$ receives a maximum of $d_i^4=d_1+d_2$ amount
of data every transfer from both $\tau_1$ and $\tau_2$. Data from a pipe
terminal is not necessarily duplicated for all pipe terminals that are
consumers. For example, $\tau_2$ generates a maximum of $d_o^2=d_2$
amount of data every transfer, by placing a single copy of the output in
a shared memory region accessible to both $\tau_4$ and $\tau_5$. If the
communication channels did not involve shared memory, then data would be
duplicated, so that $d_o^2=2d_2$.}

In Step 2, we derive a list of inequations involving period variables from the
given end-to-end timing \annotate{and scheduling} constraints in
Table~\ref{tab:example_timing}. \annotate{For simplicity, the scheduling
constraint is shown as a rate-monotonic utilization bound on the six pipe
tasks. However, for sensor inputs and some actuator outputs, our system would
map those tasks to I/O VCPUs that have a different utilization bound, as
described in our earlier work~\cite{quest-vcpu}.} The derivation is based on
Equations~\ref{equ:e2e_simple_reaction} and~\ref{equ:e2e_simple_fresh}, and
the composability property of the pipe model.  According to the conditional
equations, however, every two pipes with undetermined priority can lead to two
possible inequations. This exponentially increases the search space for
feasible periods. In order to prune the search space, our strategy is to
always start with the case where $T^{p}>T^{c}$. This is based on the
observation that tasks tend to over-sample inputs for the sake of better
overall responsiveness. Thus, the reaction constraint
$E_{{\tau_2 \mapsto \pi_2}|{\tau_5 \mapsto \pi_5}|{\tau_6 \mapsto \pi_6}} \leq
25$, for example, is translated to inequation $T^{5} + C^{2} - \Delta +
T^{6} \leq 25$. This is derived by combining
Equations~\ref{equ:e2e_simple_reaction}
and \ref{equ:e2e_simple_reaction_inc}. It is then possible to translate all
timing constraints to inequations with only periods as variables. In addition,
periods are implicitly constrained by $T^{j} > C^{j}, \forall
j \in \{1,2,\cdots,n\}$.

Given all the inequations, Step 3 attempts to find the maximum value for each
period so that the total CPU utilization is minimized. We are then left with a
linear programming problem. Unfortunately, there is no polynomial time
solution to the integer linear programming problem, as it is known to be
NP-hard. Though linear programming solutions are still available under certain
mathematical conditions~\cite{bradley1977applied}, this is beyond the scope of
this paper. Instead, in practice, the problem can be simplified because 1)
there are usually a small number of fan-in and fan-out pipe ends for each
task, meaning that a period variable is usually involved in a small number of
inequations, and 2) a sensor task period is usually pre-determined by a
hardware sampling rate limit.  For example, if we assume $T^{3}$ is known to
be 5, a feasible set of periods for the example in
Table~\ref{tab:example_timing} is easily found: $\{T^{1} = 10, T^{2} = 15,
T^{3} = 10, T^{4} = 10,T^{5} = 15,T^{6} = 5\}$.  If we ignore the integer
requirement, it is possible to find a feasible solution in polynomial time
using rational numbers that are rounded to integers. Though rounding may lead
to constraint violations, it is possible to increase the time resolution to
ensure system overheads exceed those of rounding errors. In the worst case,
the designer is always able to perform an exhaustive search of all possible
constraint solutions.

\section{Evaluation}
\label{sect:eval}
This section describes both simulations and experiments on the Intel Aero
board with an Atom x7-Z8750 1.6 GHz 4-core processor and 4GB RAM.


\subsection{Simulation Experiments}
We developed simulations for both Linux and Quest, to predict the worst-case
end-to-end time using the equations in Section~\ref{sect:e2e_analysis}. Each
simulation consists of three tasks, $\tau_1, \tau_2$ and $\tau_3$, mapped to
pipes, $\pi_1, \pi_2$ and $\pi_3$, respectively.  All three tasks search for
prime numbers within a certain range and then communicate with one another to
exchange their results.  $\tau_1$ communicates with $\tau_2$, which further
communicates with $\tau_3$. The communication channel is shared memory with
caches disabled and the data size is set to 6.7 KB to achieve a non-negligible
1 millisecond communication overhead. Each task is assigned a different search
range and the profiled execution time is shown in Table~\ref{tab:exp_setting}
in milliseconds.  The budget of each pipe is set to be slightly larger than
the execution time of its corresponding task, to compensate for system
overheads. The settings of each pipe terminal (PT) are also shown in
Table~\ref{tab:exp_setting}, again in milliseconds. Apart from the
three main tasks, the system is loaded with low priority background tasks that
consume all the remaining CPU resources and serve as potential interferences.

\begin{table}[!h]
\begin{tabular}{|c|c|c|c|c|c|c|} \hline
	& \texttt{PT 1} & \texttt{PT 2} & \texttt{PT 3} & \texttt{$\tau_1$} & \texttt{$\tau_2$} & \texttt{$\tau_3$}\\ \hline
	\texttt{Case 1} & (10,50)  & (10,150) & (5,100) & 9.5 & 9.5 & 4.5 \\ \hline
	\texttt{Case 2} & (5,100) & (10,50) & (10,150) & 4.5 & 9.5 & 9.5 \\ \hline
\end{tabular}
\caption{Simulation Settings}
\label{tab:exp_setting}
\end{table}

\begin{figure}[!ht]
  \begin{floatrow}
    \ffigbox{%
      \includegraphics[scale=0.54]{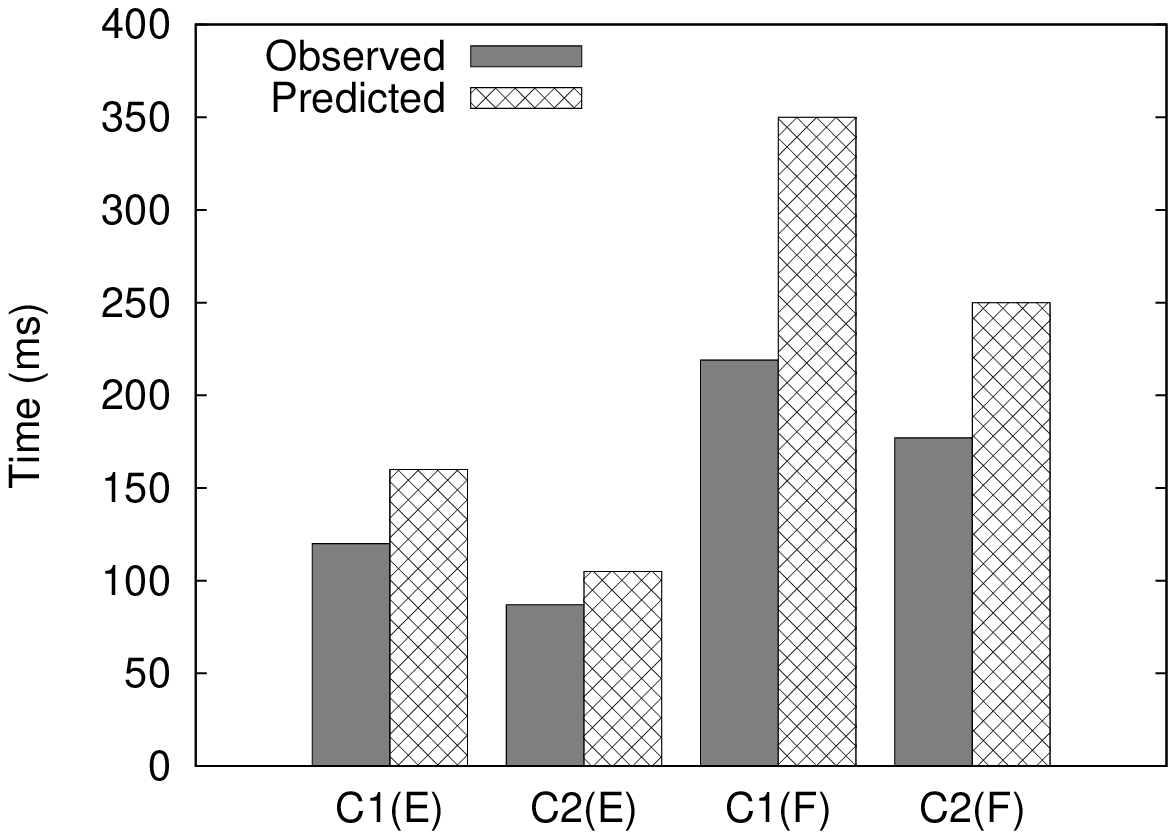}
    }{%
      \caption{Observed vs. Predicted}%
      \label{fig:observed_vs_predicted}
    }
    \ffigbox{%
      \includegraphics[scale=0.54]{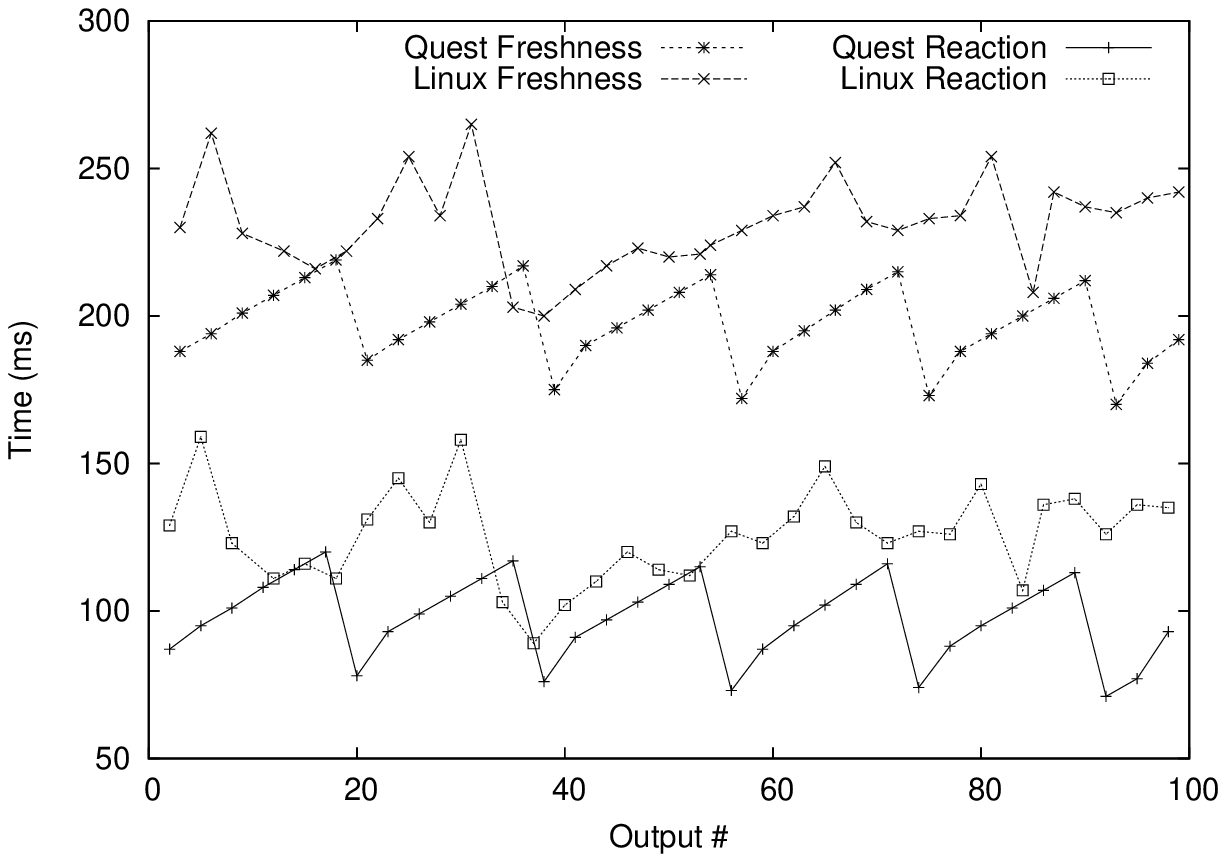}
    }{%
      \caption{Quest vs. Linux}%
      \label{fig:linux_vs_quest}
    }
  \end{floatrow}
\end{figure}

For each case, we measure the end-to-end reaction time and freshness time
separately and compare them to corresponding theoretical bounds.
Figure~\ref{fig:observed_vs_predicted} shows the results after 100,000 outputs
are produced by $\tau_3$. $C1(E)$ and $C1(F)$ are the end-to-end reaction and
freshness times, respectively, for Case 1 in Table~\ref{tab:exp_setting},
while $C2(E)$ and $C2(F)$ are for Case 2. As can be seen, the observed values
are always within the prediction bounds. As reference, we also perform the two
cases on Yocto Linux shipped with the Aero board. The kernel is version
4.4.76 and patched with the PREEMPT\_RT patch. While running the simulation,
the system also uncompresses Linux source code in the background. This places
the same load on the system as the background tasks in Quest.
Figure~\ref{fig:linux_vs_quest} shows all (not only worst case) end-to-end
reaction and freshness times within the first 100 outputs.  Compared to Linux,
there is less variance shown by the end-to-end times on Quest. Additionally,
the freshness and reaction times are generally lower with Quest than Linux.
Figure~\ref{fig:quest_vs_linux2} summarizes the worst case reaction time (WCR),
maximum variance of WCR (MaxRV), worst case freshness time (WCF) and maximum
variance of WCF (MaxFV) for both Quest and Linux.

\subsection{The Cleanflight Experiment}
\label{sect:cleanflight}
\begin{figure}[!ht]
  \centering \includegraphics[scale=0.52]{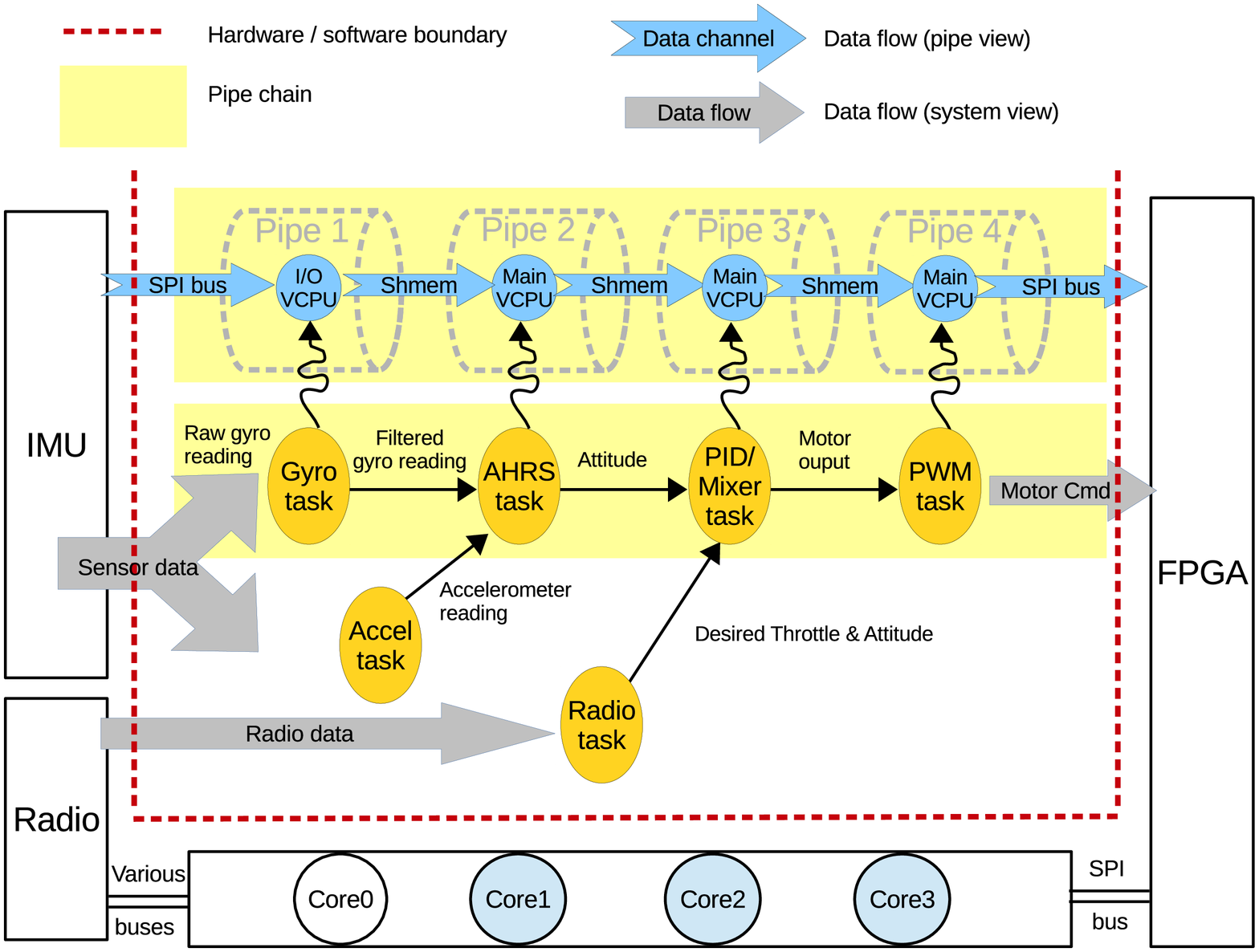}
  \caption{Cleanflight Data Flow}
  \label{fig:arch}
\end{figure}

Our next experiments apply the end-to-end design approach to determining the
periods of each task in the re-implementation of Cleanflight.  We decouple
software components in the original Cleanflight firmware and re-build the
flight controller as a multithreaded application running on the Intel Aero
board.  The hardware and software architecture is shown in
Figure~\ref{fig:arch}.


\textbf{Hardware.} We currently only use Core 0 to run Cleanflight on
Quest. The remaining three cores are reserved for our Quest-V separation
kernel~\cite{Quest-v} to run a general purpose OS such as Linux.  Apart from
the main processor, the Aero board also has an FPGA-based I/O coprocessor.  It
provides FPGA-emulated I/O interfaces including analog-to-digital conversion
(ADC), UART serial, and pulse-width modulation (PWM).  Our system currently
uses the I/O hub to send PWM signals to electronic speed controllers (ESCs)
that alter motor and, hence, rotor speeds of the drone. We modified the FPGA
logic to improve the timing resolution of PWM signals, as well as control
their duty cycle and periods. We make additional use of an onboard Bosch
BMI160 Inertial Measurement Unit (IMU).  Both the I/O hub and the IMU are
connected to the main processor via SPI bus.

\textbf{Software.} To minimize the engineering efforts, we currently disable
the auxiliary features of Cleanflight such as telemetry, blackbox data
logging, and UART-based flight control configuration. The essential components
are shown as circular tasks in Figure~\ref{fig:arch}.  The AHRS sensor fusion
task takes the input readings of the accelerometer and gyroscope (in the
BMI160 IMU) and calculates the current attitude of the drone. Then, the PID
task compares the calculated and target attitudes, and feeds the difference to
the PID control logic. In the original Cleanflight code, the target attitude
is determined by radio-control signals from a human flying the drone. In an
autonomous setting, the target attitude would be calculated according to
on-board computations based on mission objectives and flight conditions. The
output is nonetheless mixed with the desired throttle and read in by the PWM
task, which translates it to motor commands. Motor commands are sent over the
SPI bus and ultimately delivered as PWM signals to each ESC associated with a
separate motor-rotor pair on the multirotor drone.  We decouple all these
tasks into separate threads. For safety reasons, the sensor tasks and the PWM
task are given individual address spaces.  To simplify the experiment setup,
we instrument Cleanflight to use a synthetic radio input value (20\% throttle
and 0 degree pitch, roll, and yaw angle) instead of reading from the real
radio driver.  The tasks are profiled and execution times are shown in
Table~\ref{tab:cf_exec}.

\begin{table}[h]
\begin{tabular}{|c|c|c|c|c|c|c|c|} \hline
	& \texttt{Gyro} & \texttt{AHRS} & \texttt{PID} & \texttt{PWM} & \texttt{Accl} & \texttt{Radio} \\ \hline
	\texttt{Exec Times ($\mu$s)} & 174 & 10 & 2 & 970 & 167 & 12 \\ \hline
\end{tabular}
\caption{Task Execution Times}
\label{tab:cf_exec}
\end{table}


As can be seen in Figure~\ref{fig:arch}, there are currently three data paths,
originating from the gyro, the accelerometer and the radio receiver,
respectively. Unfortunately, there is little information available on what
end-to-end timing constraints should be imposed on each path to guarantee a
working drone. Most timing parameters in the original Cleanflight are
determined by trial and error. On the other hand, instead of determining the
optimum timing constraints, the focus of this paper is on guaranteeing given
constraints. Therefore, we first port Cleanflight to Yocto Linux on the Aero
board, as a reference implementation. The Linux version remains
single-threaded and used to estimate the desired end-to-end time.  For
example, for the gyro path, the worst case reaction and freshness times are
measured to be 9769 and 22972 $\mu$s, respectively.  We round them to 10 and
23 ms, and use them as end-to-end timing constraints for the Quest flight
controller implementation. Using the same approach, we determine end-to-end
reaction and freshness times for the accelerometer path, which are set to 10
and 23 ms, respectively. Finally, for the radio path, we set the end-to-end
reaction and freshness times to be 20 and 44 ms, respectively.

Using the execution times in Table~\ref{tab:cf_exec} and timing constraints
above, we apply the end-to-end design approach to derive the periods.  The
results for each task are shown in Table~\ref{tab:cf_periods}.

\begin{table}[!htb]
\begin{tabular}{|c|c|c|c|c|c|c|c|} \hline
	& \texttt{Gyro} & \texttt{AHRS} & \texttt{PID} & \texttt{PWM} & \texttt{Accl} & \texttt{Radio} \\ \hline
	\texttt{Budget/Period ($\mu$s)} & 200/1000 & 100/5000  & 100/2000 & 1000/5000 & 200/1000 & 100/10000 \\ \hline
\end{tabular}
\caption{Task Periods}
\label{tab:cf_periods}
\end{table}

\begin{figure}[!ht]
  \begin{floatrow}
    \ffigbox{%
      \includegraphics[scale=0.54]{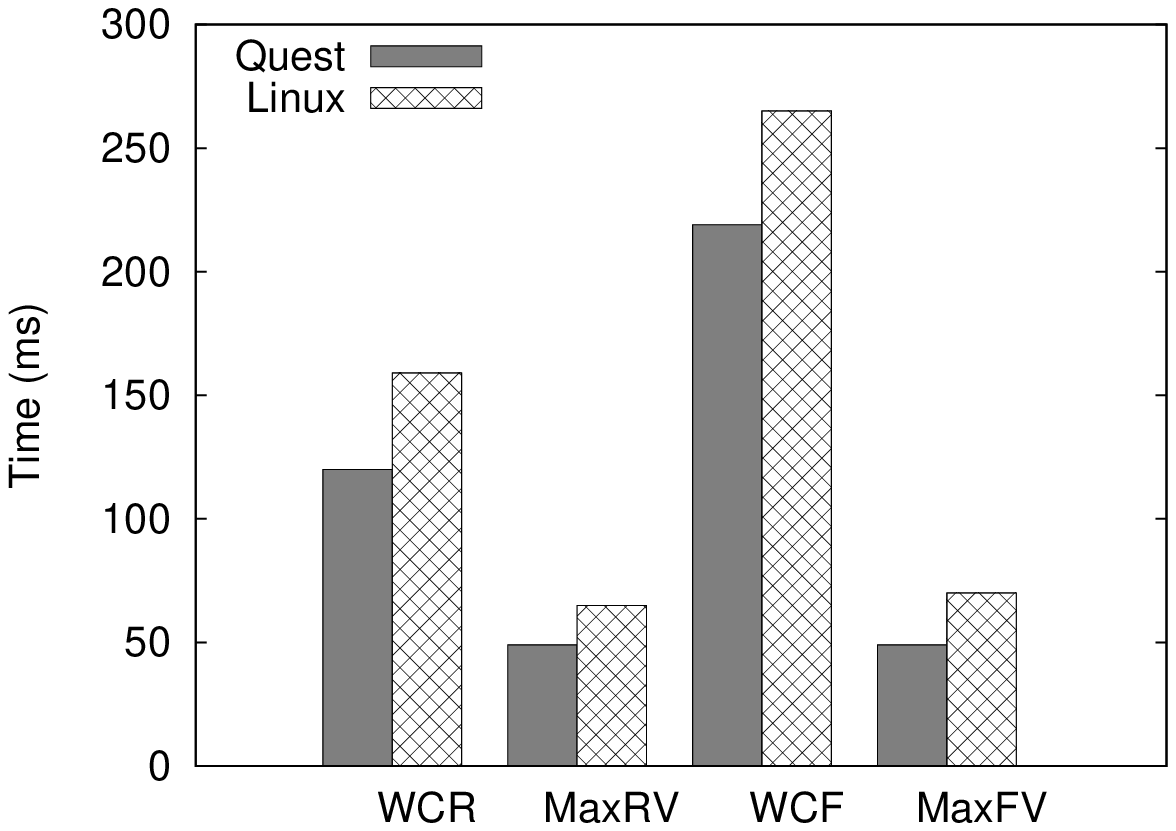}
    }{%
			\caption{Quest vs. Linux Worst-Case Times}%
			\label{fig:quest_vs_linux2}
    }
    \ffigbox{%
      \includegraphics[scale=0.54]{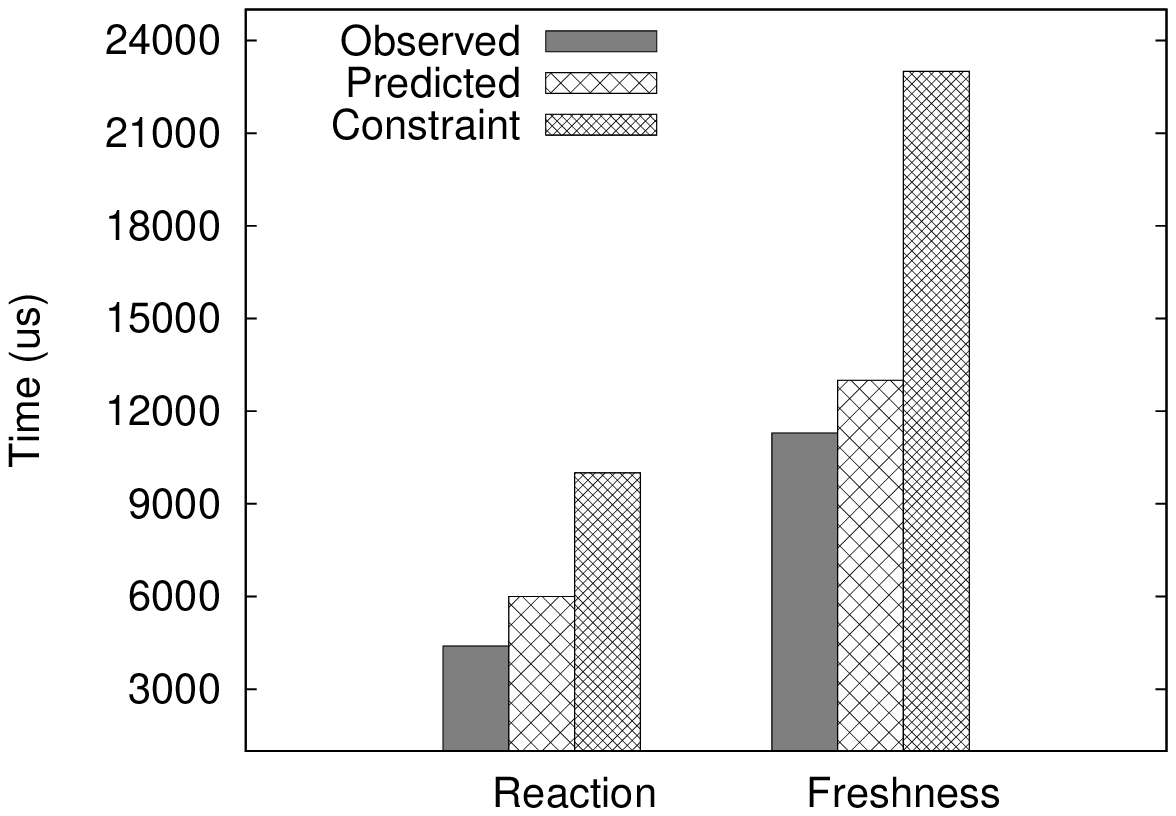}
    }{%
			\caption{Cleanflight Times}
      \label{fig:cf_exp}
    }
  \end{floatrow}
\end{figure}

\textbf{Evaluation.} To measure the actual end-to-end time, we focus on the
longest pipe chain highlighted in Figure~\ref{fig:arch}. We instrument the
Cleanflight code to append every gyro reading with an incrementing ID, and
also record a timestamp before the gyro input is read. The timestamp is then
stored in an array indexed by the ID. Every task is further instrumented to
maintain the ID when translating input data to output. This way, the ID is
preserved along the pipe chain, from the input gyro reading to the output
motor command.  After the PWM task sends out motor commands, it looks up the
timestamp using its ID and compares it to the current time. By doing this, we
are able to log both the reaction and freshness end-to-end time for every
input gyro reading.  We then compare the observed end-to-end time with the
given timing constraints, as well as the predicted worst case value.  Results
are shown in Figure~\ref{fig:cf_exp}. As can be seen, the observed values are
always within the predicted bounds, and always meet the timing constraints.

\section{Related Work}
\label{sect:related}
Feiertag \textit{et al.}~\cite{Feiertag2008} distinguish four semantics of
end-to-end time and provides a generic framework to determine all the valid
data paths for each semantic. The authors do not perform timing analysis as no
scheduling model is assumed.  Hamann \textit{et al.} also discuss end-to-end
reaction and age time~\cite {ecrts2017comm}. Their work focuses on integrating
three different communication models, including the implicit communication
model, into existing timing analysis tools such as
SymTA/S~\cite{SymTAS}. While our composable pipe model is also based on
implicit communication, we perform timing analysis using the Quest RTOS's task
and scheduling model. A large portion of end-to-end reaction time analysis is
based on the synchronous data-flow graph (SDFG)~\cite{SDFG} where inter-task
communication is driven by the arrival of input data. In recent work~\cite
{SanjoyDataFlow2017}, Singh \textit{et al.}  enhance the standard SDFG to
allow the specification of an end-to-end latency constraint.  

Gerber \textit{et al.}~\cite{Gerber1995} propose a synthesis approach that
determines tasks' periods, offsets and deadlines from end-to-end timing
constraints. \annotate{Their work relies on task precedence constraints as
there is no scheduling model used for the analysis. Our work uses a scheduling
model based on Quest to perform end-to-end timing analysis. We then derive
task periods and budgets to ensure specific reaction, freshness and
schedulability constraints.}

There are also efforts to develop programming languages, such as
Prelude~\cite{Pagetti2011} and Giotto~\cite{Giotto}, which are able to derive
tasks' periods based on user-specified timing constraints.  Kirsch~\textit{et
al.}~\cite {Giotto_helicopter} use Giotto to reimplement a helicopter control
system. Others have developed data-triggered rather than time-triggered drone
flight control using reactive programming languages~\cite{reactive-drone}.

In general operating systems, Scout~\cite{scout} exposes paths that are
similar to pipe chains in our model, to offer Quality of Service guarantees to
applications.  Paths in Scout are non-preemptive schedulable entities ordered
according to an EDF policy. 

Lastly, programming environments such as ROS~\cite {ROS}, OROCOS~\cite{orocos}
and LCM~\cite{LCM}, already widely adopted in the design of robotics and
autonomous cars, are changing the development of cyber-physical
applications. Publisher-subscriber paradigms used by ROS, for example, have
influenced our thinking in the design cyber-physical systems with modularity
and robustness amongst software components. We aim to augment Cleanflight
functionality with ROS-style services as part of our ongoing efforts to build
autonomous drones.




\section{Conclusions and Future Work}
\label{sect:conclusion}
In this paper, we identify two semantics of end-to-end time, namely reaction
and freshness time. We analyze them in the context of the Quest RTOS, for a
port of the well-known Cleanflight flight control firmware implemented as a
collection of user-level tasks. This paper describes a composable pipe model
that is built on task, scheduling and communication abstractions. Using the
pipe model, we derive the worst case end-to-end time for data flow through a
chain of tasks under various conditions.  We argue that end-to-end timing
properties should be factored in at the early stage of application design.
\annotate{Thus, we provide a mathematical framework to derive feasible task
  periods and budgets using the Quest scheduling framework to satisfy a given
  set of end-to-end timing and schedulability constraints.}  We demonstrate
the applicability of our design approach by using it to port the Cleanflight
flight controller to Quest on the Intel Aero board.

Future work will integrate the end-to-end design approach into our
Quest-Arduino (Qduino~\cite{Qduino}) development environment. This will
provide the basis for the design and implementation of an autonomous multicore
flight management system on our Quest-V separation
kernel~\cite{Quest-v}. Quest-V enables non-time-critical tasks to run on a
legacy Linux system in parallel with real-time tasks running on our Quest
RTOS.






\newpage
\bibliography{paper}

\begin{thebibliography}{10}

\bibitem{nvidia_jetson}
{Nvidia Jetson Board}, March 2017.
\newblock http://www.nvidia.com/object/embedded-systems.html.

\bibitem{snapdragon_flight_kit}
{Qualcomm Snapdragon Flight Kit}, March 2017.
\newblock
  https://www.intrinsyc.com/vertical-development-platforms/qualcomm-snapdragon-flight/.

\bibitem{AmazonAir}
Amazon~Prime Air.
\newblock http://www.amazon.com/b?ie=UTF8{\&}node=8037720011.

\bibitem{bradley1977applied}
S.P. Bradley, A.C. Hax, and T.L. Magnanti.
\newblock {\em Applied Mathematical Programming}.
\newblock Addison-Wesley Publishing Company, 1977.

\bibitem{reactive-drone}
Endri Bregu, Nicola Casamassima, Daniel Cantoni, Luca Mottola, and Kamin
  Whitehouse.
\newblock {Reactive Control of Autonomous Drones}.
\newblock In {\em Proceedings of the 14th Annual International Conference on
  Mobile Systems, Applications, and Services}, MobiSys '16, pages 207--219, New
  York, NY, USA, 2016. ACM.

\bibitem{orocos}
H.~Bruyninckx.
\newblock {Open Robot Control Software: the OROCOS Project}.
\newblock In {\em Proceedings 2001 ICRA. IEEE International Conference on
  Robotics and Automation (Cat. No.01CH37164)}, volume~3, pages 2523--2528
  vol.3, 2001.

\bibitem{Qduino}
Zhuoqun Cheng, Ye~Li, and Richard West.
\newblock {Qduino: A Multithreaded Arduino System for Embedded Computing}.
\newblock In {\em Proceedings of the 2015 IEEE Real-Time Systems Symposium
  (RTSS)}, RTSS '15, pages 261--272, Washington, DC, USA, 2015. IEEE Computer
  Society.

\bibitem{cleanflight}
{Cleanflight}: http://cleanflight.com/.

\bibitem{quest-vcpu}
Matthew Danish, Ye~Li, and Richard West.
\newblock {Virtual-CPU Scheduling in the Quest Operating System}.
\newblock In {\em Proceedings of the 17th Real-Time and Embedded Technology and
  Applications Symposium}, 2011.

\bibitem{response_time}
Robert~I. Davis, Sebastian Altmeyer, Leandro Indrusiak, Claire Maiza, Vincent
  Nelis, and Jan Reineke.
\newblock {An Extensible Framework for Multicore Response Time Analysis}.
\newblock {\em Real-Time Systems}, 2017.

\bibitem{Feiertag2008}
Nico Feiertag, Kai Richter, Johan Nordlander, and Jan Jonsson.
\newblock {A Compositional Framework for End-to-End Path Delay Calculation of
  Automotive Systems under Different Path Semantics}.
\newblock In {\em Proceedings of the IEEE Real-Time System Symposium - Workshop
  on Compositional Theory and Technology for Real-Time Embedded Systems,
  Barcelona, Spain}, November 30, 2008.

\bibitem{Gerber1995}
Richard Gerber, Seongsoo Hong, and Manas Saksena.
\newblock {Guaranteeing Real-Time Requirements With Resource-Based Calibration
  of Periodic Processes}.
\newblock {\em IEEE Trans. Softw. Eng.}, July 1995.

\bibitem{ecrts2017comm}
Arne Hamann, Dakshina Dasari, Simon Kramer, Michael Pressler, and Falk Wurst.
\newblock {Communication Centric Design in Complex Automotive Embedded
  Systems}.
\newblock In {\em 29th Euromicro Conference on Real-Time Systems (ECRTS 2017)},
  Leibniz International Proceedings in Informatics (LIPIcs), Dagstuhl, Germany,
  2017.

\bibitem{SymTAS}
Rafik Henia, Arne Hamann, Marek Jersak, Razvan Racu, Kai Richter, and Rolf
  Ernst.
\newblock {System Level Performance Analysis - the SymTA/S Approach}.
\newblock In {\em IEEE Proceedings Computers and Digital Techniques}, 2005.

\bibitem{Giotto}
Thomas~A. Henzinger, Benjamin Horowitz, and Christoph~Meyer Kirsch.
\newblock {\em Giotto: A Time-Triggered Language for Embedded Programming},
  pages 166--184.
\newblock Springer Berlin Heidelberg, Berlin, Heidelberg, 2001.

\bibitem{LCM}
Albert~S. Huang, Edwin Olson, and David Moore.
\newblock {LCM: Lightweight Communications and Marshalling}.
\newblock In {\em Proceedings of the IEEE/RSJ International Conference on
  Intelligent Robots and Systems (IROS), Taipei}, Oct 2010.

\bibitem{sensor_fusion}
Jonathan Kelly and Gaurav~S Sukhatme.
\newblock {Visual-Inertial Sensor Fusion: Localization, Mapping and
  Sensor-to-Sensor Self-calibration}.
\newblock {\em The International Journal of Robotics Research}, 30(1):56--79,
  2011.

\bibitem{Giotto_helicopter}
Christoph~M. Kirsch, Marco A.~A. Sanvido, Thomas~A. Henzinger, and Wolfgang
  Pree.
\newblock {\em A Giotto-Based Helicopter Control System}, pages 46--60.
\newblock Springer Berlin Heidelberg, Berlin, Heidelberg, 2002.

\bibitem{SDFG}
E.~A. Lee and D.~G. Messerschmitt.
\newblock Synchronous data flow.
\newblock {\em Proceedings of the IEEE}, 75(9):1235--1245, Sept 1987.

\bibitem{Lehoczky:89}
John Lehoczky, Lui Sha, and Ye~Ding.
\newblock {The Rate Monotonic Scheduling Algorithm: Exact Characterization and
  Average Case Behavior}.
\newblock In {\em Proceedings of the IEEE Real-Time Systems Symposium (RTSS)},
  1989.

\bibitem{RMS}
C.~L. Liu and James~W. Layland.
\newblock {Scheduling Algorithms for Multiprogramming in a Hard Real-Time
  Environment}.
\newblock {\em Journal of the ACM}, 20(1):46--61, 1973.

\bibitem{madgwick}
Sebastian~O.H. Madgwick.
\newblock {An Efficient Orientation Filter for Inertial and Inertial/Magnetic
  Sensor Arrays}.
\newblock Technical report, University of Bristol, 2010.

\bibitem{mahony}
Robert Mahony, Tarek Hamel, and Jean-Michel Pflimlin.
\newblock {Nonlinear Complementary Filters on the Special Orthogonal Group}.
\newblock {\em IEEE Transactions on Automatic Control}, 53, June 2008.

\bibitem{processorcapacity}
Clifford~W. Mercer, Stefan Savage, and Hideyuki Tokuda.
\newblock {Processor Capacity Reserves for Multimedia Operating Systems}.
\newblock Technical report, Pittsburgh, PA, USA, 1993.

\bibitem{MissimerPIBS}
E.~Missimer, K.~Missimer, and R.~West.
\newblock {Mixed-Criticality Scheduling with I/O}.
\newblock In {\em 28th Euromicro Conference on Real-Time Systems (ECRTS)},
  pages 120--130, July 2016.

\bibitem{scout}
David Mosberger and Larry~L. Peterson.
\newblock {Making Paths Explicit in the Scout Operating System}.
\newblock In {\em Proceedings of the Second USENIX Symposium on Operating
  Systems Design and Implementation}, OSDI '96, pages 153--167, New York, NY,
  USA, 1996. ACM.

\bibitem{BBC:SearchRescue}
BBC News.
\newblock {Disaster Drones: How Robot Teams can Help in a Crisis}.
\newblock goo.gl/6efliV.

\bibitem{Fohler}
R.~S. Oliver and G.~Fohler.
\newblock {Probabilistic Estimation of End-to-end Path Latency in Wireless
  Sensor Networks}.
\newblock In {\em 2009 IEEE 6th International Conference on Mobile Adhoc and
  Sensor Systems}, pages 423--431, Oct 2009.

\bibitem{Pagetti2011}
Claire Pagetti, Julien Forget, Fr{\'e}d{\'e}ric Boniol, Mikel Cordovilla, and
  David Lesens.
\newblock {Multi-task Implementation of Multi-periodic Synchronous Programs}.
\newblock {\em Discrete Event Dynamic Systems}, Sep 2011.

\bibitem{Fisher}
Bo~Peng, Nathan Fisher, and Thidapat Chantem.
\newblock {MILP-based Deadline Assignment for End-to-End Flows in Distributed
  Real-Time Systems}.
\newblock In {\em Proceedings of the 24th International Conference on Real-Time
  Networks and Systems}, RTNS '16, pages 13--22, New York, NY, USA, 2016. ACM.

\bibitem{quest}
{Quest RTOS}: http://questos.org/.

\bibitem{ROS}
Morgan Quigley, Ken Conley, Brian~P. Gerkey, Josh Faust, Tully Foote, Jeremy
  Leibs, Rob Wheeler, and Andrew~Y. Ng.
\newblock {ROS: An Open-source Robot Operating System}.
\newblock In {\em ICRA Workshop on Open Source Software}, 2009.

\bibitem{CAN_response}
S.~Quinton, T.~T. Bone, J.~Hennig, M.~Neukirchner, M.~Negrean, and R.~Ernst.
\newblock {Typical Worst Case Response-time Analysis and its use in Automotive
  Network Design}.
\newblock In {\em 2014 51st ACM/EDAC/IEEE Design Automation Conference (DAC)},
  pages 1--6, June 2014.

\bibitem{AFDX}
J.~L. Scharbarg, F.~Ridouard, and C.~Fraboul.
\newblock {A Probabilistic Analysis of End-To-End Delays on an AFDX Avionic
  Network}.
\newblock {\em IEEE Transactions on Industrial Informatics}, 5(1):38--49, Feb
  2009.

\bibitem{DJI}
Da-Jiang~Innovations Science and Technology Co.
\newblock {DJI}.
\newblock http://dji.com/.

\bibitem{4slots}
H.R. Simpson.
\newblock {Four-slot Fully Asynchronous Communication Mechanism}.
\newblock {\em IEEE Computers and Digital Techniques}, 137:17--30, January
  1990.

\bibitem{SanjoyDataFlow2017}
Abhishek Singh, Pontus Ekberg, and Sanjoy Baruah.
\newblock {Applying Real-Time Scheduling Theory to the Synchronous Data Flow
  Model of Computation}.
\newblock In {\em 29th Euromicro Conference on Real-Time Systems (ECRTS 2017)},
  Leibniz International Proceedings in Informatics (LIPIcs), Dagstuhl, Germany,
  2017.

\bibitem{spruntsporadic}
B.~Sprunt.
\newblock {Scheduling Sporadic and Aperiodic Events in a Hard Real-Time
  System}.
\newblock Technical Report CMU/SEI-89-TR-011, Software Engineering Institute,
  Carnegie Mellon, 1989.

\bibitem{Valavanis:book08}
K.~P. Valavanis.
\newblock {\em {Advances in Unmanned Aerial Vehicles}}.
\newblock Springer Science and Business Media, 2008.

\bibitem{Goddard}
Y.~Wang, M.~C. Vuran, and S.~Goddard.
\newblock {Cross-Layer Analysis of the End-to-End Delay Distribution in
  Wireless Sensor Networks}.
\newblock In {\em 2009 30th IEEE Real-Time Systems Symposium}, pages 138--147,
  Dec 2009.

\bibitem{Quest-v}
Richard West, Ye~Li, Eric Missimer, and Matthew Danish.
\newblock {A Virtualized Separation Kernel for Mixed-Criticality Systems}.
\newblock {\em ACM Trans. Comput. Syst.}, 34(3):8:1--8:41, June 2016.

\bibitem{CAN_response2}
P.~M. Yomsi, D.~Bertrand, N.~Navet, and R.~I. Davis.
\newblock {Controller Area Network (CAN): Response Time Analysis with Offsets}.
\newblock In {\em 2012 9th IEEE International Workshop on Factory Communication
  Systems}, pages 43--52, May 2012.

\end{thebibliography}

\end{document}